\documentclass{article}
\usepackage{graphicx}
\usepackage{amsmath}
\usepackage{booktabs}
\usepackage{breqn}
\usepackage{stackrel}
\usepackage[utf8]{inputenc}
\usepackage{geometry}
\begin{document}

\title{Temperature profile of an assemblage of non--isothermic linear energy converters}

\author{S. Gonzalez--Hernandez$^{1}$ and L. A. Arias--Hernandez$^{2}$ \\
$^{1,2}$Departamento de F\'{\i}sica, Escuela Superior de F\'{\i}sica y Matem\'aticas, Instituto\\
Polit\'ecnico Nacional, U. P. Zacatenco, Edif. \#9 2o piso, Ciudad de M\'exico, 07738, M\'EXICO \\
$^{1}$samayoa126@hotmail.com, $^{2}$larias@esfm.ipn.mx}

\maketitle

\begin{abstract}
In this paper, a proposal is presented to determine the temperature
profile obtained for an assemblage of non--isothermal linear
energy converters (ANLEC) published by Jim\'enez de Cisneros and Calvo
Hern\'andez \cite{Jimenez2007,Jimenez2008}. This is done without solving the Riccati's
differential equation, needed by these authors to get the temperature profile. Instead of use Riccati's equation, we deduce a first order ordinary differential equation, through the introduction of the force ratio $x_{D,I}$ of an ANLEC's machine--element which operates at some optimal regime. Additionally, we used the integration constant, that comes from the solution of this differential equation, to deduce the general heat fluxes of the ANLEC and tuning the assamblage's operation as direct energy converter or inverse energy converter. The temperature profile will serve to obtain the energetic behavior of a non--isothermal energy converter as heat engine, cooler or heat pump.\vspace{0.3cm}\newline 05.20.–y Classical statistical mechanics; 05.70-Ln Nonequilibrium and irreversible thermodynamics; 84.60.Bk Performance characteristics of energy conversion system; figure of merit.
\end{abstract}

\maketitle

\section{Introduction}
With the objective of perform non--isothermal system analysis
in the context of the linear irreversible thermodynamics (LIT), recently Jim\'enez and Calvo \cite{Jimenez2007,Jimenez2008} generalized the work of van der Broeck \cite{van05} on heat engines and Jim\'enez et al on refrigerators \cite{jimenez06}. They made a general construction of an ANLEC, based on this model its possible to arrive at the deduction of a differential equation for the profile of forces (temperature gradient) which is in terms of the fluxes $J$ and $v$. The differential equation obtained is a Riccati's equation \cite{Elsgoltz}.This equation is peculiar, it needs a particular solution to be fully resolved, this is an inconvenient since it implies the knowledge of extra information about system. In
\cite{Jimenez2007,Jimenez2008} they obtain this particular solution by making the analysis of the coupling coefficient $q$ \cite{Caplan1983}.

In this work, starting from the coupled chain proposed by Jim\'enez et al \cite{Jimenez2007,Jimenez2008},
we  deduced the fluxes coming from the general description
and introducing through them the so--called force ratio $x_{i}$ \cite{STUCKI1980}
(henceforth $i=D,\;I$). With this force ratio it is possible to find a first--order differential equation, which is integrated in an immediate way, because this equation is no longer a Riccati's equation. The additional information that is necessary to solve this differential equation (integration constant) is obtained through the knowledge of the optimal points (optimal force ratio) of the different objective functions that describe the energetics of a machine--element of the ANLEC.

Now, we will introduce the concept of direct or inverse linear energy converter (D-LEC or I-LEC) through the phenomenological Onsager equations written as follows:
\begin{equation}
\left[\begin{array}{c}
\frac{J_{1}}{\sqrt{L_{11}}}\\
\frac{J_{2}}{\sqrt{L_{22}}}
\end{array}\right]=\left[\begin{array}{cc}
\sqrt{L_{11}} & q\sqrt{L_{22}}\\
q\sqrt{L_{11}} & \sqrt{L_{22}}
\end{array}\right]\left[\begin{array}{c}
X_{1}\\
X_{2}
\end{array}\right],\label{egons}
\end{equation}
where \cite{Caplan1983,Arias-Hernandez2008}
\begin{equation}
q^{2}\equiv\frac{L_{12}^{2}}{L_{11}L_{22}} \in \left[0,1\right].\label{q}
\end{equation}
This coefficient comes from the second law of thermodynamics and measures the degree of coupling between the fluxes. When this coefficient goes to zero the crossed effects vanish, and therefore the fluxes become independent one of the other. When $q\rightarrow1$, the relationship between the fluxes tends to a fixed mechanistic stoichiometry one. This condition is known as ideal coupling
\cite{Caplan1983}.

Now, from the entropy production of one machine--element of the ANLEC we have
\begin{equation}
\sigma=J_{1}X_{1}+J_{2}X_{2}>0,\label{s}
\end{equation}
from this equation we can establish the relation, $\left|J_{2}\,X_{2}\right|>\left|J_{1}\,X_{1}\right|,$
with $J_{1}\,X_{1}<0$ and $J_{2}\,X_{2}>0$, according to the definition
of the driven and driver fluxes respectively. Then, we can associate
the first term of the entropy production to an energy output (by temperature
unit and time), and the second to an energy input (by temperature unit and time). Then,
using these terms we can build an ANLEC's element. This machine--element is a nonzero entropy production and a nonzero power
output converter. Using the results by Caplan and Essig \cite{Caplan1983},
it is possible to advance towards a linear description of this element. In addition, we can take into account a parameter which measures the relation between the two forces: $X_{1}$, associated with the driven flux $J_{1}$, and $X_{2}$, associated with the driver flux $J_{2}$, as follows: \begin{equation}
x\equiv\sqrt{\frac{L_{11}}{L_{22}}}\frac{X_{1}}{X_{2}},\label{x}
\end{equation}
where $x\in\left[-1,0\right]$ is called the force ratio. 

\subsection{Direct ANLEC's machine--element (heat engine)}

First, we can use the entropy production of a Direct ANLEC's machine--element, given by Eq. (\ref{s}), with the heat flux $J_{D2}$ promoted by a temperature gradient $X_{D2}$, as the driver flux, and the driven flux $J_{D1}$ any other flux against $X_{D1}$, in this case
\begin{equation}
x_{D}=\sqrt{\frac{L_{11}}{L_{22}}}\frac{X_{D1}}{X_{D2}}.\label{xd}
\end{equation}
Now, by using the $q$ and $x_{D}$ parameters we can rewrite the fluxes $J_{D1}$ and $J_{D2}$ as follows:
\begin{equation}
J_{D1}=\left(1+\frac{x_{D}}{q}\right)L_{12}X_{D2},\label{jd1}
\end{equation}
and
\begin{equation}
J_{D2}=\left(1+qx_{D}\right)L_{22}X_{D2}.\label{jd2}
\end{equation}

\subsection{Inverse ANELC's machine--element (refrigerator and heat pump)}

Now we will write the fluxes $J_{I1}$ and $J_{I2}$ in terms of the inverse force ratio and the coupling coefficient for an inverse ANLEC's machine--element, by using the entropy production Eq.(\ref{s}).In this case to use the force ratio introduced by Stucki \cite{STUCKI1980}
it is necessary to take into account that the driven flux is the heat flux, $J_{I2}$, and the driver flux (any other kind of flux) is $J_{I1}$ which are associated with the forces $X_{I2}$ and $X_{I1}$ respectively, then the inverse force ratio $x_{I}$, such that $x_{I}\in\left[-1,0\right]$, is as follows:
\begin{equation}
x_{I}=\sqrt{\frac{L_{22}}{L_{11}}}\frac{X_{I2}}{X_{I1}}.\label{xi}
\end{equation}
Then the fluxes $J_{I1}$ and $J_{I2}$ Eq.(\ref{egons})
in terms of the inverse force ratio $x_{I}$ and the coupling coefficient $q$ are
\begin{equation}
J_{I1}=\left(1+\frac{1}{qx_{I}}\right)L_{12}X_{I2},\label{ji1}
\end{equation}
and
\begin{equation}
J_{I2}=\left(1+\frac{q}{x_{I}}\right)L_{22}X_{I2}.\label{ji2}
\end{equation}

This paper is organized as follows: in Section \ref{assemblage}
the ANLEC and the considerations made by \cite{Jimenez2007,Jimenez2008}
for this construction are presented. In Section \ref{force} the force ratio
$x_{i}$ is introduced, this is done through the generalized fluxes
and forces $J_{1}$ and $J_{2}$, which leads to a first-order equation
whose integration is immediate except for an integration constant.
In Subsection \ref{coefficient} the integration
constant is determined considering the strong coupling limit and small
temperature differences. In Section \ref{dlec} several objective functions
for the Direct ANLEC's element are presented, besides we present the approximation of these
objective functions for the case when the difference of temperature
is small. In the Section \ref{ilec} several objective
functions that describe the energetics of the Inverse ANLEC's element, in the subsection
\ref{refrigerator} is presented the system operated as a
refrigerator and in subsection \ref{bomb} is presented
the system operated as a heat pump, finally, the conclusions are presented
in section \ref{conclusions}.

\section{Assemblage of a non\textendash isothermal linear energy converters\label{assemblage}}

\begin{figure}[tb]
\centering a)\includegraphics[width=0.2\textwidth]{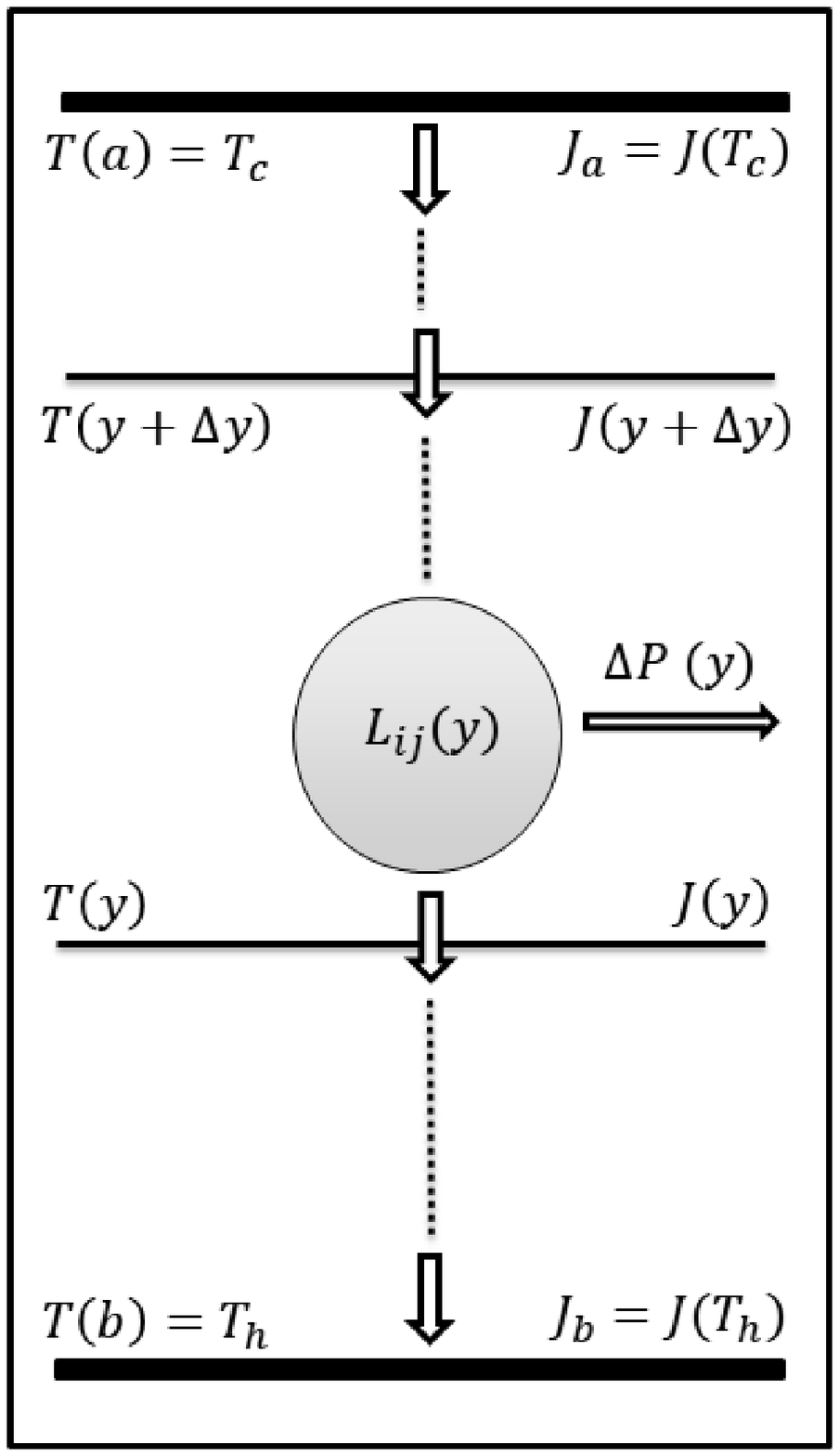} b)\includegraphics[width=0.2\textwidth]{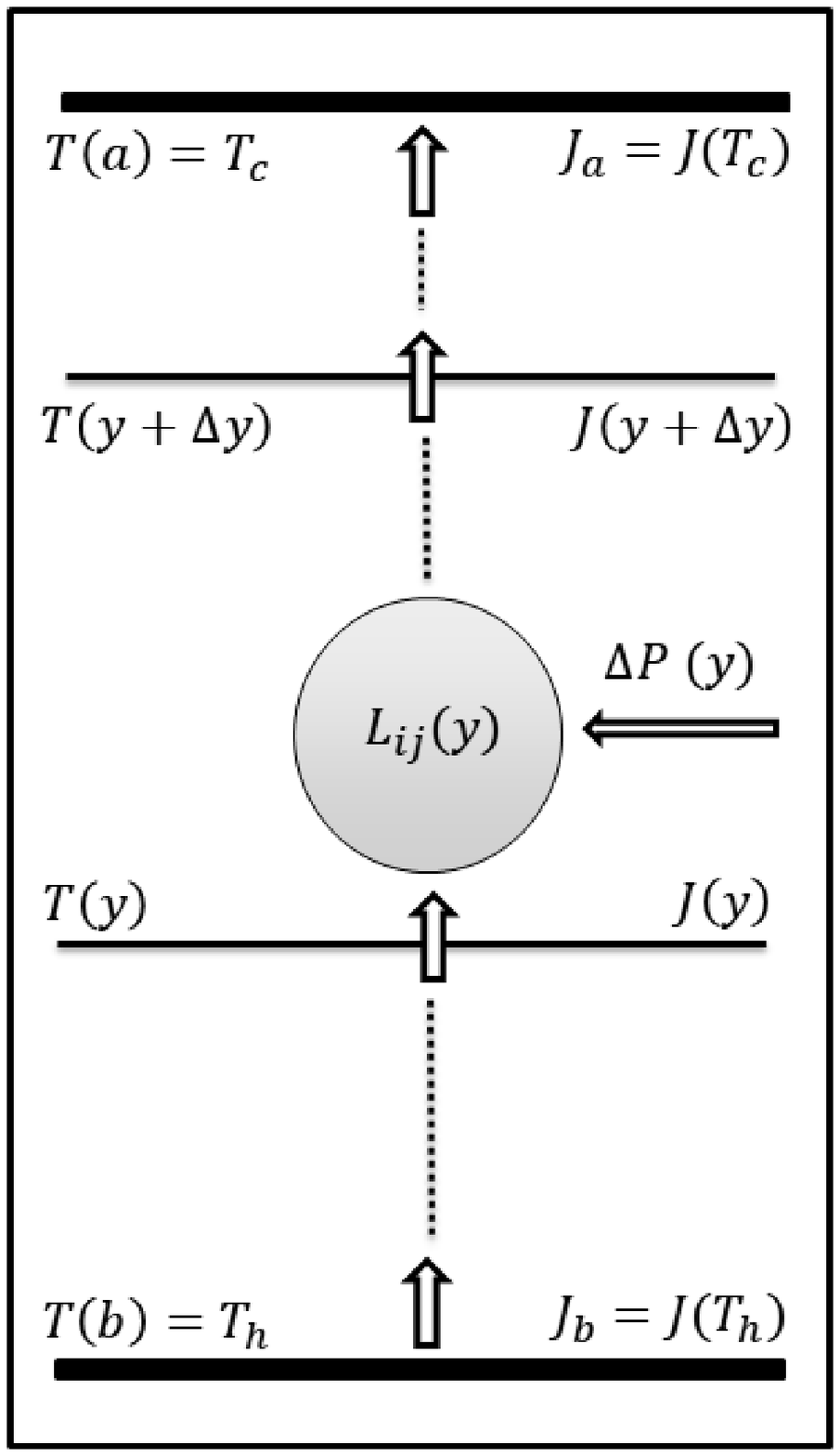} 
\begin{quote}
\caption[Arreglo de maquinas termicas acopladas.]{ The figure a) shows ANLEC operating directly (engine) where we can
see that the arrows represent the direction in which the assembly
moves a heat load from the hot reservoir $T(b)=T_{h}$ to cold reservoir
$T(a)=T_{c}$, in this process the device delivers a part of this
heat load which is transformed by the system into a useful work $\delta P\left(y\right)$.
In the same way, the figure b) shows ANLEC operating in the opposite
way (cooler and heater). Here we can observe how the arrows represent
the sense in which a heat load is extracted from the cold reservoir
$T(a)=T_{c}$ to the hot reservoir $T(b)=T_{h}$, to achieve it is
necessary to enter a work quantity $\delta P\left(y\right)$.}

{\footnotesize{}\label{Fig1} }{\footnotesize \par}
\end{quote}
\end{figure}

The starting point of this work is the assemblage of a non--isothermal
linear energy converters described by \cite{Jimenez2007,Jimenez2008},
for which it is necessary to do the following construction: we will
consider that each assemblage of converters works between two reservoirs
at different fixed temperatures, and each assemblage member is a non-isothermal
linear energy converter. This converter works between a $\Delta T=T\left(y+\Delta y\right)-T\left(y\right)\neq0$
temperature difference, where $T\left(y+\Delta y\right)$ is the low
temperature of the previous converter and $T\left(y\right)$ is the
high temperature of the next converter (following the heat flux direction
we can construct a D\textendash LEC or I\textendash LEC). These reservoirs
are labeled by the y-coordinate ($y\in\left[a,\;b\right]$) and form
a temperature profile $T(y)$ which varies from $T(b)=T_{h}$ to $T(a)=T_{c}$.
In addition, it must be considered that the all converters operate
in a stationary state. On the other hand, the converters are individually
coupled in the following sense: the input or (output) of heat per
unit of time (or per period) of the device unit $T(y+\triangle y)$,
is exactly equal to the output or (input) heat on the next device
$T(y)$. Therefore, the entire assemblage can be considered as a single
energy converter whose overall behavior is determined by the heat
exchange with the deposits in $y=a$ and $y=b$, as shown in the generic
device Fig. (\ref{Fig1}).
 Now, let the heat $J\left(y+\triangle y\right)$ and $J(y)$ flow
to the reservoirs $T(y+\triangle y)$ and $T(y)$, respectively, the
difference between these flows is exchanged with the environment as
the work per unit of time $dW/dt=\overset{.}{W}=P$ to a rate $\delta P\left(y\right)=f\left(y\right)v\left(y\right)\triangle y$.
This work is done against an external force $f\left(y\right)\triangle y$
and $v\left(y\right)$ is a conjugate flux$\delta P\left(y\right)=f\left(y\right)v\left(y\right)\triangle y$.
The conservation of energy in this system implies that $J\left(y+\triangle y\right)=J\left(y\right)+f\left(y\right)v\left(y\right)\triangle y$,
taking the limit $\triangle y\rightarrow0$ we obtain

\begin{equation}
J'\left(y\right)=f\left(y\right)v\left(y\right),\label{dj}
\end{equation}
where $J'\left(y\right)=dJ(y)/dy$.
The integration of this equation results in the net power delivered
by the total assemblage in terms of the heat fluxes $J_{a}=J\left(T_{h}\right)$
and $J_{b}=J\left(T_{c}\right)$:

\begin{equation}
P=\int_{a}^{b}dyf\left(y\right)v\left(y\right)=J_{b}-J_{a}.\label{p}
\end{equation}
The entropy production rate of the system  operating between the reservoirs $T(y)$ and $T(y+\Delta y),$ can be written as

\begin{equation}
\Delta\overset{.}{S}\left(y\right)=\frac{J\left(y+\triangle y\right)}{T\left(y+\Delta y\right)}-\frac{J\left(y\right)}{T\left(y\right)}.\label{delta-s}
\end{equation}
Taking into account the conservation of energy Eq.(\ref{dj}), we
obtain at first order in $\Delta y$

\begin{equation}
\Delta\overset{.}{S}\left(y\right)=\left[\frac{J'\left(y\right)}{T(y)}-\frac{J\left(y\right)}{T(y)^{2}}\right]\Delta y=\left[\frac{f\left(y\right)v\left(y\right)}{T(y)}-\frac{J\left(y\right)}{T(y)^{2}}\right]\Delta y.\label{ent}
\end{equation}
 this last expression implies that in the limit when $lim_{\Delta y\rightarrow0}\left[\Delta\overset{.}{S}\left(y\right)/\Delta y\right]=\overset{.}{S}'\left(y\right)=\left[J'\left(y\right)/T(y)\right]-\left[J\left(y\right)/T(y)^{2}\right]=\left[J\left(y\right)/T(y)\right]'$.
The integration of $a$ to $b$ results in the total entropy production rate

\begin{equation}
\overset{.}{S}=\sigma=\frac{J_{b}}{T_{h}}-\frac{J_{a}}{T_{c}}.\label{sent}
\end{equation}
From the analysis of the entropy production Eq.(\ref{ent}) we should
consider $\left[1/T(y)\right]'=-T(y)'/T^{2}\left(y\right)$
and $f\left(y\right)/T(y)$ as thermodynamic forces with conjugate
fluxes $J\left(y\right)$ and $v\left(y\right)$ \cite{Yourgrau} respectively,
under the considerations of the LIT \cite{Yourgrau,degroot84} the
phenomenological Onsager's equations Eq.(\ref{egons}) can be written
as follows:

\begin{equation}
v\left(y\right)=L_{11}\left(y\right)\frac{f\left(y\right)}{T\left(y\right)}+L_{12}\left(y\right)\left[\frac{1}{T\left(y\right)}\right]^{'},\label{f1}
\end{equation}

\begin{equation}
J\left(y\right)=L_{21}\left(y\right)\frac{f\left(y\right)}{T\left(y\right)}+L_{22}\left(y\right)\left[\frac{1}{T\left(y\right)}\right]^{'},\label{f2}
\end{equation}
It should be noted that once the temperatures $T_{h}$, $T_{c}$ and
the Onsager's coefficients $L_{ij}(y)$ are set, the profile strength
$f(y)$ can not be freely chosen. The reason is that the conservation
of the energy in each motor expressed by Eq.(\ref{dj}), together
with the dynamic equations Eq.(\ref{f1}) and Eq.(\ref{f2}) implies
an Riccati differential equation for $f(y)$ \cite{Elsgoltz}:

\begin{equation}
\left[L_{12}\left(y\right)f\left(y\right)\right]^{'}=L_{11}\left(y\right)f\left(y\right){}^{2}+T\left(y\right)\left\{ L_{22}\left(y\right)\left[\frac{1}{T\left(y\right)}\right]^{'}\right\} ^{'}.\label{eq:11-1-1}
\end{equation}

\section{Force ratio for the assemblage of a non\textendash isothermal
linear energy converter\label{force}}

Starting from the equations Eq.(\ref{jd1}-b) and Eq.(\ref{jd2}-a)
for the case when the system is operating as a D-LEC, and the equations
Eq.(\ref{ji1}-b) and Eq.(\ref{ji2}-a) when the system is operating
as a cooler or heat pump (I-LEC), and rewrite the dynamic equations
Eq.(\ref{f1}) and Eq.(\ref{f2}), for which we should only identify
$J_{1}=v\left(y\right)$, $J_{2}=J\left(y\right)$ and $X_{1}=f\left(y\right)/T\left(y\right)$,
$X_{2}=\left[1/T\left(y\right)\right]^{'}$ whereby the equations
Eq.(\ref{f1}) and Eq.(\ref{f2}) can be written as follows

\begin{equation}
v\left(y\right)=A_{i}\left(y\right)L_{12}\left(y\right)\left[\frac{1}{T\left(y\right)}\right]^{'},\label{f11}
\end{equation}

\begin{equation}
J\left(y\right)=B_{i}\left(y\right)L_{12}\left(y\right)\frac{f\left(y\right)}{T\left(y\right)},\label{f22}
\end{equation}
where the functions $A_{i}\left(y\right)$ and $B_{i}\left(y\right)$,
if we talk about the system operated as a heat engine (direct converter)
then $(A_{D}\left(y\right),\;B_{D}\left(y\right)$), or cooler or
heat pump (inverse converter) then ($A_{I}\left(y\right),\;B_{I}\left(y\right)$),
now replacing Eq.(\ref{f11}) and Eq.(\ref{f22}) at Eq.(\ref{dj})
we obtain the following differential equation

\begin{equation}
\left[B\left(y\right)\frac{f\left(y\right)}{T\left(y\right)}L_{12}\left(y\right)\right]^{'}=f\left(y\right)L_{12}\left(y\right)\left[\frac{1}{T\left(y\right)}\right]^{'}A\left(y\right),\label{deq1}
\end{equation}
or

\begin{equation}
\left[B\left(y\right)f\left(y\right)L_{12}\left(y\right)\right]^{'}=T\left(y\right)f\left(y\right)L_{12}\left(y\right)\left[\frac{1}{T\left(y\right)}\right]^{'}\left[A\left(y\right)-B\left(y\right)\right],\label{deq2}
\end{equation}
where we can consider that

\begin{equation}
A_{D}\left(y\right)=1+x_{D}\left(y\right)/q\left(y\right)\quad and\quad B_{D}\left(y\right)=1+1/\left[q\left(y\right)x_{D}\left(y\right)\right],\label{ADyBD}
\end{equation}
for the heat engine case, and

\begin{equation}
A_{I}(y)=1+1/\left[q\left(y\right)x_{I}\left(y\right)\right]\quad and\quad B_{I}(y)=1+x_{I}\left(y\right)/q\left(y\right),\label{AIyBI}
\end{equation}
for the case of refrigerator or heat pump, in principle $q\left(y\right)$
and $x\left[q\left(y\right)\right]$, for simplicity we consider that
$x\left[q\left(y\right)\right]=x\left(q\right)$, $A_{i}(y)=A_{i}$,
$B_{i}(y)=B_{i}$. The consideration that $x\left[q\left(y\right)\right]=x\left(q\right)$
is quite reasonable since the optimal values of various modes of operation
depend only of the coupling coefficient $q$. The optimal values
of the direct force ratio $x_{D}$ and inverse force ratio $x_{I}$,
in this context, for different modes of operation $q$, Eq.(\ref{deq2})
takes the following form

\begin{equation}
Z\left(y\right)'=\omega_{i}Z\left(y\right)T\left(y\right)\left[\frac{1}{T\left(y\right)}\right]^{'},\label{deqz}
\end{equation}
where

\begin{equation}
\omega_{i}=\left(A_{i}-B_{i}\right)/B_{i},\label{wi}
\end{equation}
and $Z\left(y\right)=f\left(y\right)L_{12}$, with this we can write
Eq.(\ref{deqz}) as

\begin{equation}
\frac{dZ\left(y\right)}{Z\left(y\right)}=-\omega_{i}\frac{1}{T\left(y\right)}dT\left(y\right),\label{deqzbis}
\end{equation}
integrating we obtain,

\begin{equation}
Z\left(y\right)=c_{i}T\left(y\right){}^{-\omega_{i}},\label{z}
\end{equation}
with $c_{i}$ is an integration constant to be determined, returning
to the variable $f(y)$ we have

\begin{equation}
f(y)=\frac{c_{i}}{L_{12}\left(y\right)T\left(y\right){}^{\omega_{i}}},\label{perf}
\end{equation}
finally substituting Eq.(\ref{perf}) at Eq.(\ref{f22}) we get

\begin{equation}
J\left(y\right)=c_{i}\frac{B_{i}\left(y\right)}{T\left(y\right){}^{\alpha_{i}}},\label{perj}
\end{equation}
where 

\begin{equation}
\alpha_{i}=\omega_{i}+1.\label{alpha_i}
\end{equation}

\subsection{Determination of the integration constant \label{coefficient}}

In order to determine completely the temperature profile Eq.(\ref{perf})
it is necessary to determine the integration constant to small difference
of temperatures $c_{0_{i}}$, where $i=I,D$. In principle, considering
a general treatment we do not have this condition, but we can make
some considerations to be able to find the solution when certain particular
requirements are met. Consider the equations Eq.(\ref{f1}) and Eq.(\ref{f2})
when the coupling coefficient $q=1$, it happens that $v\left(y\right)=0\rightarrow J\left(y\right)=0$,
take them Eq.(\ref{f1}) and equal zero and solve for $f\left(y\right)$,
likewise we can take Eq.( \ref{f2}) and match to zero and solving
in the same way for $f\left(y\right)$, then we have a particular
solution $f_{0}\left(y\right)$

\begin{equation}
f_{0}\left(y\right)=-T\left(y\right)\frac{L_{12}}{L_{11}}\left[\frac{1}{T\left(y\right)}\right]^{'}=-T\left(y\right)\frac{L_{22}}{L_{12}}\left[\frac{1}{T\left(y\right)}\right]^{'},\:\longrightarrow L_{12}^{2}=L_{22}L_{11}\longrightarrow q=1.\label{q1}
\end{equation}
Equalizing Eq.(\ref{perf}) and Eq.(\ref{q1}), and solve to find
the value of the constant $c=c_{o_{i}}$

\begin{equation}
c_{0_{i}}=-q^{2}L_{22}T\left(y\right)^{\alpha_{i}}X_{2}.\label{cq1}
\end{equation}
Now we will consider Eq.(\ref{f11}) and let $v\left(y\right)=0$,
we find $A\left(y\right)L_{12}\left[1/T\left(y\right)\right]^{'}=0\:\longrightarrow A\left(y\right)=0$,
since $L_{12}\left(1/T\left(y\right)\right)^{'}\neq0$, so for a heat
engine: $A_{D}(y)=1+x_{D}\left(y\right)/q\left(y\right)=0$, then we find

\begin{equation}
A_{D}(y)=0\rightarrow x_{D}=-q.\label{ad1}
\end{equation}
Now if we substitute Eq.(\ref{ad1}) at Eq.(\ref{cq1}), we obtain
an integration constant for the D\textendash LEC

\begin{equation}
c_{0_{D}}=qx_{D}L_{22}T\left(y\right)^{\alpha_{D}}X_{2}.\label{cd}
\end{equation}
On the other hand, consider the case of the inverse converter $A_{I}\left(y\right)=1+1/\left[q\left(y\right)x_{I}\left(y\right)\right]=0$, then

\begin{equation}
A_{I}\left(y\right)=0\rightarrow x_{I}=-\frac{1}{q},\label{ai1}
\end{equation}
whereby we substitute Eq.(\ref{ai1}) at Eq.(\ref{cq1}), obtaining
an integration constant for the I\textendash LEC

\begin{equation}
c_{0_{I}}=\frac{q}{x_{I}}L_{22}T\left(y\right)^{\alpha_{I}}X_{2}.\label{ci}
\end{equation}
We must note that this constant is only valid for values of $q=1$ (perfect coupling) and for small temperature differences, notice that Eq.(\ref{ad1}) and Eq.(\ref{ai1}) agree with the values of the force ratio when the system is operated at minimum dissipation function, in another hand,  we must mention that  the integration constant that we find is valid for a special case in which the  temperature difference   is small, so, this constant only can be used after to make this approximation.

\section{Assemblage as heat engine (Direct Linear Energy Converter D-LEC)\label{dlec}}

\subsection{Heat engine}

The heat engines are thermal machines that exchanging an amount of energy with the surroundings to do power output.   In  this  case  a  gradient  of temperature  promotes  a  flux  against  any  other  gradient  (gravity,  electric  field,  etc.). Some models of this
kind of engines have been proposed in the context of Linear Irreversible
Thermodynamics, Finite\textendash Time Thermodynamics and other constructions
within Non\textendash Equilibrium Thermodynamics \cite{van05,jimenez06,Arias-Hernandez2008}.
On the other hand, one of the most important features of an irreversible
converter, is the amount of energy exchanged with the surroundings
to do work or acomplish another type of objective. This feature is
usually known as the energetics of  LEC \cite{Arias-Hernandez2008}.
We can write some functions that characterize this energetics in terms of the fluxes $J_{a}$ and $J_{b}$, some of which we present in the following section.

\subsection{Efficiency}

We can define a measure of performance of an energy converter

\begin{equation}
r\equiv\frac{Useful\:Energy}{Input\:Energy}=\frac{\varepsilon_{u}}{\varepsilon_{in}}.\label{r}
\end{equation}
 This  quantity  measures  the performance  of energy  conversion, depending of the energy
conversion objective of the converter , in such a way that if we operate
the converter as a heat engine D\textendash LEC this paremeter is
known as the efficiency $\eta$, and in the case of the refrigerator
and heat pump these performance measures are known as cooling Coefficient
of Performance (COP) $\epsilon$ and heating COP ${\epsilon}_{{H}}$,
respectively. For a thermal engine, the power output (useful energy)
is given by $P=J_{b}-J_{a}$ Eq.(\ref{pj}), then the efficiency is given by

\begin{equation}
\eta=\frac{P}{J_{b}}=1-\tau^{-\alpha_{D}(x_{D},q)},\label{eq:28a}
\end{equation}
where $\tau=\left(T_{c}/T_{h}\right)\in\left[0,1\right]$, the results
for different values of the force ratio $x_{D}$ and the alpha value $\alpha_{D}(x_{D},q)$ can shown in  Tab.(\ref{tab:2}). In the
same way we can calculate the maximum efficiency $\eta_{M\eta}$,
for which we perform the optimization of Eq.(\ref{eq:28a}) with respect
to $x_{D}$, and solving the equation $\left.\partial\eta/\partial x_{D}\right|_{x_{DM\eta}}=0$,
whereby we find the optimal value of the force ratio $x_{DM\eta}$

\begin{equation}
x_{DM\eta}=-\frac{q}{1+\sqrt{1-q^{2}}},\label{xmn}
\end{equation}
now, substituting the equation Eq.(\ref{xmn}) at Eq.(\ref{eq:28a})
we obtain the maximum value of the efficiency $\eta_{M\eta}$

\begin{equation}
\eta_{M}\left(q,\,\tau\right)=1-\tau{}^{-\alpha_{D}(x_{DM\eta},q)}.\label{eq:nmaxima}
\end{equation}
the alpha value $\alpha_{D}(x_{DM\eta},q)$ can shown in Tab.(\ref{tab:2}).

\begin{table}[tb]
\centering{\footnotesize{}}%
\begin{tabular}{cccc}
\toprule 
{\footnotesize{}$x_{D}$} & {\footnotesize{}$\alpha_{D}(x_{D},q)=\frac{x_{D}^{2}+qx_{D}}{qx_{D}+1}$} & {\footnotesize{}$\alpha_{D}(x_{D}(q=1))$} & {\footnotesize{}$\eta(x_{D}(q=1))$ }\tabularnewline
\midrule
\midrule 
{\footnotesize{}$x_{Dmfd}=-q$ } & {\footnotesize{}$0$} & {\footnotesize{}$0$} & {\footnotesize{}$0$}\tabularnewline
\midrule 
{\footnotesize{}$x_{DM\eta}=-\frac{q}{1+\sqrt{1-q^{2}}}$} & {\footnotesize{}$-\left(\frac{q}{1+\sqrt{1-q^{2}}}\right)^{2}$} & {\footnotesize{}$-1$} & {\footnotesize{}$1-\frac{T_{c}}{T_{h}}=\eta_{C}$}\tabularnewline
\midrule 
{\footnotesize{}$x_{DMP}=-\frac{q}{2}$ } & {\footnotesize{}$-\frac{q^{2}}{2\left(2-q^{2}\right)}$} & {\footnotesize{}$-\frac{1}{2}$} & {\footnotesize{}$1-\sqrt{\frac{T_{c}}{T_{h}}}=\eta_{CA}$}\tabularnewline
\midrule 
{\footnotesize{}$x_{DMFE}=-\frac{3}{4}q$} & {\footnotesize{}$-\frac{3}{4}\left(\frac{q^{2}}{4-3q^{2}}\right)$} & {\footnotesize{}$-\frac{3}{4}$} & {\footnotesize{}$1-\left(\frac{T_{c}}{T_{h}}\right)^{\frac{3}{4}}$}\tabularnewline
\midrule 
{\footnotesize{}$x_{DM\Omega_{G}}=x_{DME_{G}}=-\frac{q^{2}\left(q^{2}-4\right)}{4\left(q^{2}-2\right)}$} & {\footnotesize{}$-\frac{q^{3}\left(q^{2}-4\right)\left\{ q\left[q\left(q-4)-4\right)\right]+8\right\} }{4\left(q^{2}-2\right)\left\{ q^{2}\left[q^{3}-4\left(q+1\right)\right]+8\right\} }$} & {\footnotesize{}$-\frac{3}{4}$} & {\footnotesize{}$1-\left(\frac{T_{c}}{T_{h}}\right)^{\frac{3}{4}}$}\tabularnewline
\bottomrule
\end{tabular} \caption{In the table the different values of $\alpha_{D}$ are shown for different
operation modes, besides the evaluation of this $\alpha$ in the efficiency
Eq.(\ref{eq:28a}) considering the perfect coupling value $q=1$,
it is possible to observe that when $\eta(x_{DM\eta}(q=1))$ the efficiency
 obtained is that of Carnot (C) $\eta_{C}$ \cite{carnot}, likewise
when $\eta(x_{DMP}(q=1))$, the efficiency of Curzon and Ahlborn (CA)
is obtained $\eta_{CA}$ \cite{carnot,CA}.}

\label{tab:2} 
\end{table}
On the other hand we can consider the case of small\textbf{ $\triangle T$},
we can approximate first order Eq.(\ref{eq:28a}) as follows

\begin{equation}
\eta=1-\left(\frac{T}{T+\Delta T}\right)^{-\alpha_{D}}\approx-\alpha_{D}\frac{\Delta T}{T}=-\alpha_{D}\eta_{c},\label{eq:33a}
\end{equation}
where $\eta_{c}=\left(\Delta T/T\right)$ is the efficiency of carnot,
whereby efficiency for small\textbf{ $\triangle T$} can be written
as follows

\begin{equation}
\eta(x,\alpha_{D},\eta_{c})=-\left(\frac{x_{D}^{2}+qx_{D}}{qx_{D}+1}\right)\eta_{c},\label{eq:30a}
\end{equation}
 the value $\alpha_{D}(x_{D},q)$ and their respective efficiency
are shown in Tab.(\ref{tab:2}). 
\subsection{Power output}

Consider the case when the system operates as D-LEC, now sustitute
$\alpha_{D}$ in Eq.(\ref{p}) thus the power output $P=J_{b}-J_{a}$ Eq.(\ref{pj}) can be written as follows

\begin{equation}
P=J_{b}-J_{a}=c_{D}\left(1+\frac{1}{qx_{D}}\right)\left[\frac{1}{T_{h}{}^{\alpha_{D}\left(x_{D},q\right)}}-\frac{1}{T_{c}{}^{\alpha_{D}\left(x_{D},q\right)}}\right],\label{pj}
\end{equation}
For small temperatures differences such that $T_{c}=T$ and $T_{h}=T+\Delta T$,
we can approximate to first order Eq.(\ref{pj}) for small $\Delta T$
as follows

\begin{equation}
\left(\frac{1}{T+\Delta T}\right)^{\alpha_{D}\left(x_{D},q\right)}\approx\frac{1}{T^{\alpha_{D}\left(x_{D},q\right)}}-\alpha_{D}\left(x_{D},q\right)\frac{\Delta T}{T}\frac{1}{T^{\alpha_{D}\left(x_{D},q\right)}}=\frac{1}{T^{\alpha_{D}\left(x_{D},q\right)}}\left[1-\alpha_{D}\left(x_{D},q\right)\eta_{C}\right],\label{td}
\end{equation}

replace Eq.(\ref{cd}) and Eq.(\ref{td}) at Eq.(\ref{pj}), obtaining

\begin{equation}
P=-\eta_{c}x_{D}\left(x_{D}+q\right)L_{22}X_{2}.\label{eq:27a-2}
\end{equation}

\subsection{Efficient power}

The efficient power $P_{\eta}$ is defined as $P_{\eta}=P\eta$ \cite{Yilmaz},
which can be written as follows

\begin{equation}
P_{\eta}=c_{D}\left(1+\frac{1}{qx_{D}}\right)\left[\frac{1}{T_{h}^{\alpha_{D}\left(x_{D},q\right)}}-\frac{1}{T_{c}^{\alpha_{D}\left(x_{D},q\right)}}\right]\left[1-\left(\frac{T_{h}}{T_{c}}\right)^{\alpha_{D}\left(x_{D},q\right)}\right].\label{eq:41}
\end{equation}
Consider the case of small\textbf{ $\triangle T$}, substitute temperatures
Eq.(\ref{td}) and Eq.(\ref{cd}) at Eq.(\ref{eq:41}), so we get

\begin{equation}
P_{\eta}=\eta_{c}^{2}\frac{\left[x\left(x+1\right)\right]}{qx+1}L_{22}X_{2}.
\end{equation}

\subsection{Dissipation function}

Before starting with the dissipation function $\Phi_{D}$, it is necessary to
write the entropy production Eq.(\ref{sent}), whereby when replacing
the fluxes $J_{b}$ and $J_{a}$ Eq.(\ref{perj}), is obtained

\begin{equation}
\sigma_{D}=c_{D}\left(1+\frac{1}{qx_{D}}\right)\left[\frac{1}{T_{c}{}^{\alpha_{D}\left(x_{D},q\right)+1}}-\frac{1}{T_{h}{}^{\alpha_{D}\left(x_{D},q\right)+1}}\right].\label{eq:43a}
\end{equation}
Consider the case of small\textbf{ $\triangle T$}, substitute temperatures
Eq.(\ref{td}) and Eq.(\ref{cd}) at Eq.(\ref{eq:43a}), so we get

\begin{equation}
\sigma_{D}=\frac{\eta_{c}}{T}\left(x^{2}+2qx_{D}+1\right)L_{22}X_{2},
\end{equation}
we can define the dissipation function as the energy that the machine
discards, which is defined by Tribus \cite{tribus61} as the entropy
production $\sigma=\dot{S}$, multiplied by the cold reservoir $T_{c}$
in the case of D-LEC $\Phi_{D}$ and by the hot reservoir $T_{h}$
for the refrigerators and heat pumps I-LEC $\Phi_{I}$:

\begin{equation}
\Phi_{D}=T_{c}\sigma_{D},\label{eq:3-1}
\end{equation}
and

\begin{equation}
\Phi_{I}=T_{h}\sigma_{I}.\label{eq:4b}
\end{equation}
From Eq.(\ref{eq:3-1}) it is possible to calculate the function of
discipation when the system operates I-LEC, for which it is necessary
to calculate the production of entropy $\sigma_{D}$ Eq.(\ref{eq:3-1}),
$\Phi_{D}=T_{c}\left[\left(J_{a}/T_{a}\right)-\left(J_{b}/T_{b}\right)\right]$,
substituting the flows we obtain

\begin{equation}
\Phi_{D}=c_{D}T_{c}\left(1+\frac{1}{qx_{D}}\right)\left[\frac{1}{T_{c}{}^{\alpha_{D}\left(x_{D},q\right)+1}}-\frac{1}{T_{h}{}^{\alpha_{D}\left(x_{D},q\right)+1}}\right],\label{eq:44b}
\end{equation}
consider the case of small\textbf{ $\triangle T$}, substituting Eq.(\ref{td})
and Eq.(\ref{cd}) at Eq.(\ref{eq:44b}), with what we obtain

\begin{equation}
\Phi_{D}=\eta_{c}\left(x_{D}^{2}+2qx_{D}+1\right)L_{22}X_{2}.
\end{equation}

\subsection{Generalized Ecological function}

The generalized Ecological function $E_{DG}$ \cite{tornez06} is
defined in the following manner $E_{DG}=P-g_{E}^{D}\left(\eta_{MP}\right)\Phi_{D}$,
where $P$ is the output power Eq.(\ref{pj}), $g_{E}^{D}\left(\eta_{MP}\right)$
is the function $g_{E}\left(\eta\right)$ \cite{Angulo-Brown2001,Arias-Hernandez2003}
evaluated in the efficiency operated at maximum power and $\Phi_{D}$
 is the dissipation function Eq.(\ref{eq:44b}), the $g_{E}$ function
is defined in the following way $g_{E}^{D}(\eta)=\eta/(\eta_{c}-\eta)$,
the evaluation of the efficiency Eq.(\ref{eq:28a}) in the ratio force
at maximum power $x_{DMP}$ Eq.(\ref{tab:2}) is 
\begin{equation}
\eta(x_{DMP}=-\frac{q}{2})=\eta_{Mp}=1-\tau^{-\alpha_{D}(x_{DMP},q)},~\label{n(xmp)}
\end{equation}
whereby we can compute $g_{E}^{D}\left(\eta_{MP}\right)=\eta_{MP}/(\eta_{c}-\eta_{MP})$, with what we obtain

\begin{equation}
g_{E}^{D}\left(\eta_{MP}\right)=\frac{1-\tau^{-\alpha_{D}(x_{DMP},q)}}{\;\tau^{-\alpha_{D}(x_{DMP},q)}-\tau},\label{eq:48a}
\end{equation}
with which we can write the generalized ecological function $E_{DG}$in
the following manner 

\begin{equation}
E_{DG}=c_{D}\left(1+\frac{1}{qx_{D}}\right)\left\{ \frac{T_{h}+g_{E}^{D}\left(\eta_{MP}\right)T_{c}}{T_{h}{}^{\alpha_{D}\left(x_{D},q\right)+1}}-\frac{\left[g_{E}^{D}\left(\eta_{MP}\right)+1\right]}{T_{c}^{\alpha_{D}\left(x_{D},q\right)}}\right\} .\label{eq:26a}
\end{equation}
Notice that when $g_{E}^{D}\left(\eta_{MP}\right)=1$ in Eq.(\ref{eq:26a})
implies that $E_{DG}=E_{D}$, where $E_{D}$ is the ecological function
\cite{angulo91}. On the other hand, consider the case of small\textbf{
$\triangle T$}, so that substituting Eq.(\ref{tab:2}) at Eq.(\ref{eq:48a})
is obtained

\begin{equation}
g_{E}^{D}\left(\eta_{MP}\right)=\frac{\eta_{Mp}}{\eta_{c}-\eta_{MP}}\approx\frac{q^{2}}{4-3q^{2}},\label{eq:51a}
\end{equation}
substituting Eq.(\ref{eq:51a}), Eq.(\ref{td}) and Eq.(\ref{cd})
at Eq.(\ref{eq:26a}) with which we obtain finally

\begin{equation}
E_{DG}=\eta_{c}\frac{x_{D}\left[4x-q\left(q^{2}-4\right)\right]+q^{2}\left(1-2x_{D}^{2}\right)}{3q^{2}-4}L_{22}X_{2}.\label{eq:32a}
\end{equation}

\subsection{Generalized Omega function}

The generalized Omega function $\Omega_{DG}$ \cite{tornez06}, this
objective function proposes a compromise between the effective useful
energy $\varepsilon_{e,u}=\varepsilon_{u}-r_{min}\varepsilon_{in}$
, and the lost useful energy $\varepsilon_{l,u}=r_{max}\varepsilon_{in}-\varepsilon_{u}$
, where $\varepsilon_{u}$ is the useful energy of the machine,
$r_{min}$  is the minimum performance coefficient, $\varepsilon_{in}$
is the input energy and $r_{max}$ is the maximum performance coefficient,
the performance coefficient is defined as Eq.(\ref{r}),
finally the generalized Omega function is defined as follows

\begin{equation}
\Omega_{DG}=\varepsilon_{e,u}-g_{\Omega}\left(\eta_{MP}\right)\varepsilon_{l,u}.\label{eq:9B}
\end{equation}
Where $g_{\Omega}^{D}\left(\eta_{MP}\right)$ is the function $g_{\Omega}^{D}\left(\eta\right)$
\cite{tornez06} for the Omega function evaluated in the efficiency
operated at the maximum output power $\eta_{MP}$ Eq.(\ref{n(xmp)}),
which is defined as follows $g_{\Omega}^{D}\left(\eta\right)=\eta/(\eta_{M\eta}-\eta)$,
the evaluation of the efficiency Eq.(\ref{eq:28a}) in the ratio force
at maximum efficiency $x_{M\eta}$ Eq.(\ref{tab:2}) is

\begin{equation}
\eta_{M\eta}=1-\tau^{-\alpha_{D}(x_{DM\eta},q)},\label{n(xmn)}
\end{equation}
we finally get 

\begin{equation}
g_{\Omega}^{D}\left(\eta_{Mp}\right)=\frac{\eta_{MP}}{\eta_{M\eta}-\eta_{MP}}=\frac{1-\tau^{-\alpha_{D}(x_{DMP},q)}}{\tau^{-\alpha_{D}(x_{DMP},q)}-\tau^{-\alpha_{D}(x_{DM\eta},q)}},\label{eq:62a}
\end{equation}
the input energy $\varepsilon_{in}=\left(1-\tau\right)J_{h}$, and
the utility energy $\varepsilon_{u}=P$ is the power output, so the
performance for the case of a direct (heat engine) converter is the
following $r=\eta/\eta_{C}$, The minimum performance coefficient
is zero $r_{min}=0$, the maximum is $r_{max}=\eta_{M\eta}/\eta_{C}$,
so that the generalized Omega function can be written as follows

\begin{equation}
\Omega_{DG}=P\left[1+g_{\Omega}^{D}\left(\eta_{MP}\right)\right]-g_{\Omega}^{D}\left(\eta_{MP}\right)r_{max}\left(1-\tau\right)J_{h},\label{eq:30a-1-1}
\end{equation}
notice that when the equation Eq.(\ref{eq:30a-1-1}) $g_{\Omega}^{D}\left(\eta_{MP}\right)=1\longrightarrow\Omega_{DG}=\Omega_{D}$,
where $\Omega_{D}$ is the omega function \cite{calvo01}, which can
finally be written as follows

\begin{multline}
\Omega_{DG}=c_{D}\left(1+\frac{1}{q x_{D}}\right)\left\{ \left[\frac{1}{T_{h}^{\alpha_{D}\left(x_{D},q\right)}}-\frac{1}{T_{c}^{\alpha_{D}\left(x_{D},q\right)}}\right]\left[1+g_{\Omega}^{D}\left(\eta_{MP}\right)\right]\right. -\\
\left. g_{\Omega}^{D}\left(\eta_{MP}\right)r_{max}\left[\frac{T_{h}-T_{c}}{T_{h}^{\alpha_{D}\left(x_{D},q\right)+1}}\right]\Bigg\}.\right.
\label{eq:64a}
\end{multline}
On the other hand, consider the following approximation for small\textbf{
$\triangle T$} as follows

\begin{equation}
\left(\frac{T_{h}-T_{c}}{T_{h}{}^{\alpha_{D}+1}}\right)=\frac{\Delta T}{\left(T+\Delta T\right)^{\alpha_{D}+1}}\approx\frac{1}{T^{\alpha_{D}}}\frac{\Delta T}{T},\label{eq:65a}
\end{equation}
substituting the values of $\eta_{MP}$ Eq.(\ref{n(xmp)}) and $\eta_{M\eta}$
Eq.(\ref{n(xmn)}) at Eq.(\ref{eq:62a}), is obtained

\begin{equation}
g_{\Omega}^{D}\left(\eta_{MP}\right)=\frac{\eta_{MP}}{\eta_{M\eta}-\eta_{MP}}\approx\left(\frac{\sqrt{1-q^{2}}+1}{\sqrt{1-q^{2}}-1}\right)^{2},\label{eq:66a}
\end{equation}
finally substituting Eq.(\ref{eq:66a}) and Eq.(\ref{eq:65a}) at
Eq.(\ref{eq:64a}) is obtained

\begin{equation}
\Omega_{DG}=\eta_{c}\frac{x_{D}\left[q\left(q^{2}-4\right)-4x_{D}\right]+q^{2}\left(2x_{D}^{2}-1\right)}{\left(\sqrt{1-q^{2}}-1\right)^{2}}L_{22}X_{2}.\label{eq:27a-1-1-1}
\end{equation}

\section{Assemblage as Cooler or Heat Pump (Inverse Linear Energy Converter
I-LEC)\label{ilec}}

\subsection{Cooler\label{refrigerator}}

As it is well known, the coolers have the objective of using a load of input power $P$ for the extraction of a cooling load $J_{c}$ from a cold reservoir at a temperature $T_{c}$ to another reservoir at temperature $T_{h}$,  whereby it is possible
to determine the useful energy $\varepsilon_{u}$, which in the case
of this system is the cooling load $J_{c}=\varepsilon_{u}$, since
moving this load is the only objective of this converter and the input
energy $\varepsilon_{in}$ is the input power $P=\varepsilon_{in}$ supplied
to move that cooling load, with these quantities we can calculate
a measure of the cooling performance

\begin{equation}
\epsilon=\frac{\varepsilon_{u}}{\varepsilon_{in}}=\frac{J_{c}}{P}.\label{eq:COP}
\end{equation}
To move forward we must have to take the pertinent considerations
in Eq.(\ref{ci}) for which we consider the case when the system operates
as a refrigerator or heat pump (inverse converter), for which $A_{I}(y)$
and $B_{I}(y)$ Eq.(\ref{AIyBI}), replace at $\omega_{I}$ Eq.(\ref{wi}),
so we take Eq.(\ref{p}) and $\alpha_{I}=\left(qx_{I}+1\right)/\left(x_{I}^{2}+qx_{I}\right)$
Eq.(\ref{alpha_i}).

\subsection{Coefficient of performance (COP)}

For the assemblage as a cooler the measure of its performance,
called COP $\epsilon$, now replacing $J_{c}$ Eq.(\ref{perj})
and $P$ Eq.(\ref{p}) at $\epsilon$ Eq.(\ref{eq:COP}) we get

\begin{equation}
\epsilon=\frac{1}{\tau{}^{\alpha_{I}(x_{I},q)}-1}.\label{cop}
\end{equation}
The results for different values of the force ratio $x_{I}$ ,
the alpha value $\alpha_{I}$ is shown in the Tab.(\ref{tab:3-1}).
In the same way we can calculate the maximum COP: $\epsilon_{M}$,
for which we perform the optimization of Eq.(\ref{cop}) with respect
to $x_{I}$, and solving the equation $\left.\partial\epsilon/\partial x_{I}\right|_{x_{M\epsilon}}=0$,
whereby we find the optimal value of the force ratio $x_{M\epsilon}$

\begin{equation}
x_{IM\epsilon}=-\frac{q}{1+\sqrt{1-q^{2}}}.\label{xmcop}
\end{equation}
Now, substituting the equation Eq.(\ref{xmcop}) at Eq.(\ref{cop})
we obtain the maximum value of the efficiency $\epsilon_{M}$ 

\begin{equation}
\epsilon_{M}\left(q,\,\tau\right)=\frac{1}{\tau^{\alpha_{I}\left(x_{IM\epsilon},q\right)}-1},\label{maxcop}
\end{equation}
where, $\alpha_{I}\left(x_{IM\epsilon},q\right)$ is the value of
$\alpha_{I}$ evaluated in $x_{IM\epsilon}$, as we can see in Tab.(\ref{tab:3-1}).
For small $\Delta T$, we can approximate the COP Eq.\textbf{(\ref{cop})},
substituting Eq.(\ref{td}) at Eq.\textbf{(\ref{cop})}, under this
consideration we can approximate $\epsilon$ as follows

\begin{equation}
\epsilon=\frac{1/T_{c}^{\alpha_{I}(x_{I},q)}}{1/T_{h}^{\alpha_{I}(x_{I},q)}-1/T_{c}{}^{\alpha_{I}(x_{I},q)}}\approx-\frac{1}{\alpha_{I}\left(x_{I},q\right)}\epsilon_{C},\label{eq:71a}
\end{equation}
where, for small $\Delta T$, $\epsilon_{C}=T/\Delta T$,
it is the Carnot COP.

\begin{table}[tb]
\centering{\tiny{}}%
\begin{tabular}{cccc}
\toprule 
{\footnotesize{}$x_{I}$} & {\tiny{}$\alpha_{I}\left(x_{I},q\right)=\frac{qx_{I}+1}{x_{I}\left(x_{I}+q\right)}$} & {\footnotesize{}$\epsilon(x,q,\epsilon_{c})$ } & {\footnotesize{}$\epsilon(x_{I},q=1,\epsilon_{c})$ }\tabularnewline
\midrule
\midrule 
{\footnotesize{}$x_{I_{mdfc}}=-\frac{1}{q}$} & {\footnotesize{}$\stackrel[x\rightarrow-q]{q\rightarrow1}{\longrightarrow}-1$} & {\footnotesize{}$\stackrel[x\rightarrow-q]{q\rightarrow1}{\longrightarrow}-1$} & {\footnotesize{}$\epsilon_{C}$}\tabularnewline
\midrule 
{\footnotesize{}$x_{M\epsilon}=x_{M\eta}$} & $-\left(\frac{1+\sqrt{1-q^{2}}}{q}\right)^{2}$ & {\footnotesize{}$-1$} & {\footnotesize{}$\epsilon_{C}$}\tabularnewline
\midrule 
{\footnotesize{}$x_{I_{ME_{GI}}}=x_{I_{MG\Omega I}}=\frac{2q}{q^{2}-4\left(1+\sqrt{1-q^{2}}\right)}$} & $\frac{\left[q^{2}-4\left(1+\sqrt{1-q^{2}}\right)\right]\left[3q^{2}-4\left(1+\sqrt{1-q^{2}}\right)\right]}{2q^{2}\left[q^{2}-2\left(1-2\sqrt{1-q^{2}}\right)\right]}$ & {\footnotesize{}$-\frac{2}{3}$} & {\footnotesize{}$\frac{2}{3}\epsilon_{C}$}\tabularnewline
\bottomrule
\end{tabular} \caption{In the table the different values of $1/\alpha_{I}\left(x_{I},q\right)$
are shown for different operation modes, in addition the evaluation
of this $1/\alpha_{I}\left(x_{I},q\right)$ in the COP Eq.(\ref{eq:71a})
(for the case of small $\Delta T$) considering the perfect coupling
value $q=1$. }

\label{tab:3-1} 
\end{table}

\subsection{Cooling dissipation function}

Then we will calculate the cooling dissipation function $\varPhi_{I}$
of the assemblage, for which it is necessary to first calculate
the production of cooling entropy $\sigma_{I}=\left(J_{h}/T_{h}\right)-\left(J_{c}/T_{c}\right)$
which after replacing Eq.(\ref{perj}) can be written as follows

\begin{equation}
\sigma_{I}=c_{I}\left(1+\frac{x_{I}}{q}\right)\left[\frac{1}{T_{h}{}^{\alpha_{I}(x_{I},q)+1}}-\frac{1}{T_{c}{}^{\alpha_{I}(x_{I},q)+1}}\right].\label{eq:24b}
\end{equation}

By considering the case of small difference of temperatures,
and substituting Eq.(\ref{td}) and Eq.(\ref{ci})
at Eq.(\ref{eq:43a}), we obtain

\begin{equation}
\sigma_{I}=-\frac{1}{\epsilon_{C}}\left(\frac{x_{I}^{2}+2qx_{I}+1}{x_{I}^{2}}\right)\frac{L_{22}X_{2}}{T}.\label{eq:74b}
\end{equation}
Now with the aid of the entropy production Eq.(\ref{eq:74b}),
we can write the dissipation function of cooling Eq.(\ref{eq:4b})

\begin{equation}
\Phi_{I}=c_{I}T_{h}\left(1+\frac{x_{I}}{q}\right)\left[\frac{1}{T_{h}{}^{\alpha_{I}(x_{I},q)+1}}-\frac{1}{T_{c}{}^{\alpha_{I}(x_{I},q)+1}}\right],\label{eq:25b}
\end{equation}
and for small $\triangle T$ we get

\begin{equation}
\Phi_{I}=-\frac{1}{\epsilon_{C}}\left(\frac{x_{I}^{2}+2qx_{I}+1}{x_{I}^{2}}\right)L_{22}X_{2}.\label{eq:25b-1}
\end{equation}

\subsection{Generalized Ecological function of cooling}

The generalized Ecological function of cooling $E_{IG}$,
for the assemblage, is defined in the following manner \cite{tornez06} 

\begin{equation}
E_{IG}=J_{c}-g_{E}^{I}\left(\frac{\epsilon_{M}}{2}\right)\Phi_{I},\label{eq:47a}
\end{equation}
where $\Phi_{I}$ is the dissipation function of cooling
Eq.(\ref{eq:24b}) and $g_{E}^{I}\left(\epsilon_{M}/2\right)$ is
the function $g_{E}^{I}(\epsilon)=\left(\epsilon_{C}\epsilon\right)/\left(\epsilon_{C}-\epsilon\right)$
\cite{Angulo-Brown2001,Arias-Hernandez2003} evaluated at the half
of the maximum COP, $\epsilon_{M}$ Eq.(\ref{maxcop}):

\begin{equation}
g_{E}^{I}\left(\frac{\epsilon_{M}}{2}\right)=\frac{\epsilon_{C}\left(\epsilon_{M}/2\right)}{\epsilon_{C}-\left(\epsilon_{M}/2\right)}=\frac{\epsilon_{C}T_{h}{}^{\alpha_{I}\left(x_{M\epsilon},q\right)}}{2\left[T_{c}{}^{\alpha_{I}\left(x_{M\epsilon},q\right)}-T_{h}{}^{\alpha_{I}\left(x_{M\epsilon},q\right)}\right]\epsilon_{C}-T_{h}{}^{\alpha_{I}\left(x_{M\epsilon},q\right)}},\label{eq:81a}
\end{equation}
with which we can write $E_{IG}$ in the following manner

\begin{equation}
E_{IG}=c_{I}\left(1+\frac{x_{I}}{q}\right)\left\{ \frac{1}{T_{c}^{\alpha_{I}\alpha_{I}(x_{I},q)}}-g_{E}^{I}\left(\frac{\epsilon_{M}}{2}\right)\left[\frac{1}{T_{c}^{\alpha_{I}(x_{I},q)+1}}-\frac{1}{T_{h}^{\alpha_{I}(x_{I},q)+1}}\right]T_{h}\right\} ,\label{eq:59a}
\end{equation}
notice that when $g_{E}^{I}\left(\epsilon_{M}/2\right)=1$ then $E_{IG}=E_{I}$,
where $E_{I}$ is the Ecological function for cooling. For small\textbf{
$\triangle T$} , we can approximate at first order the equation Eq.(\ref{eq:81a})
as follows

\begin{equation}
g_{E}^{I}\left(\epsilon_{M}\right)=\frac{\epsilon_{C}\left(\epsilon_{M}/2\right)}{\epsilon_{C}-\left(\epsilon_{M}/2\right)}\approx\frac{\epsilon_{c}q^{2}}{2\left(1+\sqrt{1-q^{2}}\right)^{2}-q^{2}},\label{eq:86a}
\end{equation}
finally when replacing Eq.(\ref{ci}), Eq.(\ref{ci}) and Eq.(\ref{eq:86a})
at Eq.(\ref{eq:59a}), is obtained

\begin{equation}
E_{IG}=\left[\frac{x_{I}+q}{x_{I}}+\left(\frac{q}{x_{I}}\right)^{2}\frac{x_{I}^{2}+2qx_{I}+1}{2\left(1+\sqrt{1-q^{2}}\right)^{2}-q^{2}}\right]L_{22}X_{2}.\label{eq:59a-1}
\end{equation}

\subsection{Generalized Omega function of cooling}

The generalized Omega function of cooling $\Omega_{IG}$ was proposed
by Tornez for an irreversible FTT\textendash Refrigerator \cite{tornez06},
this objective function proposes a compromise between the effective
useful energy $\varepsilon_{u,e}=\varepsilon_{u}-r_{min}\varepsilon_{in}$,
and the lost useful energy $\varepsilon_{l,u}=r_{max}\varepsilon_{in}-\varepsilon_{u}$
, where $\varepsilon_{u}=J_{c}$ is the useful energy of cooling,
$r_{min}=0$ is the minimum performance coefficient, $\varepsilon_{in}=P=J_{h}-J_{c}$
is the input energy and $r_{max}=\epsilon_{M}$is the maximum performance
coefficient, we can write $\Omega_{IG}$ as follows

\begin{equation}
\Omega_{IG}=\varepsilon_{u,e}-g_{\Omega}^{I}\left(\frac{\epsilon_{M}}{2}\right)\varepsilon_{l,u},\label{eq:52a-1}
\end{equation}
where the parameter $g_{\Omega}^{I}\left(\frac{\epsilon_{M}}{2}\right)$
corresponds to the function $g_{\Omega}^{I}\left(\epsilon\right)=\epsilon/\left(\epsilon_{Max}-\epsilon\right)$
\cite{tornez06} evaluated at the half of maximum COP: $g_{\Omega}^{I}\left(\frac{\epsilon_{M}}{2}\right)=1$,
following these definitions we can write $\Omega_{IG}$ as

\begin{equation}
\Omega_{IG}=2J_{c}-\epsilon_{M}P=\left(2+\epsilon_{M}\right)J_{c}-\epsilon_{M}J_{h},\label{eq:86b}
\end{equation}
finally when replacing Eq.(\ref{perj}), Eq.(\ref{eq:86b}) and Eq.(\ref{cop}),
getting

\begin{equation}
\Omega_{IG}=c_{I}\left(1+\frac{x_{I}}{q}\right)\left\{ \left(2+\epsilon_{M}\right)\frac{1}{T_{c}^{\alpha_{I}(x_{I},q)}}-\epsilon_{M}\left[\frac{1}{T_{h}^{\alpha_{I}(x_{I},q)}}\right]\right\} .\label{eq:87b}
\end{equation}
In the case of small\textbf{ $\triangle T$}, we can substitute Eq.(\ref{td})
and Eq.(\ref{ci}) at Eq.(\ref{eq:87b}) to obtain

\begin{equation}
\Omega_{IG}=\Omega_{I}=\left(\frac{q+x_{I}}{x_{I}}\right)\frac{1}{\epsilon_{C}}\left[2\epsilon_{C}+\alpha_{I}(x_{I},q)\epsilon_{M}\right]L_{22}X_{2},\label{eq:55a-1}
\end{equation}
we can note that $\Omega_{IG}=\Omega_{I}$, where $\Omega_{I}$ is
the Omega function of cooling.

\subsection{Efficient cooling power}

We can define the efficient cooling power $P_{\epsilon}$,
which is the product of the cooling power $J_{c}$ Eq.(\ref{perj}),
by the COP $\epsilon$ Eq.(\ref{eq:COP}), we can write $P_{\epsilon}=J_{c}\epsilon$
as follows  

\begin{equation}
P_{\epsilon}=c_{I}\left(1+\frac{x_{I}}{q}\right)\frac{1}{T_{c}^{\alpha_{I}(x_{I},q)}}\frac{T_{h}{}^{\alpha_{I}(x_{I},q)}}{\left[T_{c}{}^{\alpha_{I}(x_{I},q)}-T_{h}{}^{\alpha_{I}(x_{I},q)}\right]}.\label{eq:71b}
\end{equation}
In the limit of small difference of temperatures, by taking
Eq.(\ref{eq:71a}) and Eq.(\ref{ci}) and substituting
at Eq.(\ref{eq:71b}) we obtain

\begin{equation}
P_{\epsilon}=-\frac{x_{I}\left(x_{I}+q\right)}{x_{I}^{2}}\epsilon_{c}L_{22}X_{2}.
\end{equation}

\subsection{Heat pump\label{bomb}}

As is well known heat pumps are refrigeration engines that take heat
from a cold reservoir $J_{c}$ and transfer it to a hotter one $J_{h}$
thanks to a external power $P$, $i.e.$ does exactly the same as
refrigerators, what sets them apart is the energy conversion objective.
In refrigerators the objective is to cool and keep the cold reservoir
at low temperature, and in heat pumps the objective is to provide
heat and keep the hot reservoir at high temperature, we can note that
since the heat pump and the refrigerator are actually the same engine,
but with different energy conversion objective, the entropy output
of the heat pump and the refrigerator are the same Eq.(\ref{eq:24b}),
and in addition given the approximation of small diference
of temperatures $T_{c}\approx T_{h}\approx T$ also the dissipation
function of cooling $\Phi_{I}$ and the dissipation function of heating,
$\Phi_{IH}=T_{h}\sigma$, will be the same $\Phi_{I}=\Phi_{IH}=T\sigma$.

\subsection{Heating COP}

The heating COP $\epsilon_{H}$ for the assemblage is defined in the following
manner $\epsilon_{H}=J_{h}/P$, wherewith we can replace  $J_{h}$ Eq.(\ref{perj})
and $P$ Eq.(\ref{p}) at $\epsilon_{H}$, we get

\begin{equation}
\epsilon_{H}=\frac{T_{c}{}^{\alpha_{I}(x_{I},q)}}{T_{c}{}^{\alpha_{I}(x_{I},q)}-T_{h}{}^{\alpha_{I}(x_{I},q)}}=\frac{\tau^{\alpha_{I}(x_{I},q)}}{\tau^{\alpha_{I}(x_{I},q)}-1}.\label{copH}
\end{equation}
the results for different values of the force ratio $x_{I}$,
$\alpha_{I}$ and their respective COP are shown in Tab (\ref{tab:3-1},).
In the same way we can calculate the maximum COP $\epsilon_{MH}$,
for which we perform the optimization of (\ref{cop}) with respect
to $x_{I}$, and solving the equation $\left.\partial\epsilon_{H}/\partial x_{I}\right|_{x_{MH}}=0,$
whereby we find the optimal value of the force ratio $x_{MH}$

\begin{equation}
x_{MH}=-\frac{q}{1+\sqrt{1-q^{2}}},\label{xcopH}
\end{equation}
now, substituting the equation Eq.(\ref{xcopH}) at Eq.(\ref{copH})
we obtain the maximum value of the heating COP $\epsilon_{MH}$

\begin{equation}
\epsilon_{MH}\left(q,\,\tau\right)=\frac{\tau^{\alpha_{I}(x_{MH},q)}}{\tau^{\alpha_{I}(x_{MH},q)}-1}.\label{maxcopH}
\end{equation}
consider the case of small $\Delta T$, substituting Eq.(\ref{td})
at Eq.(\ref{copH}), under this consideration we can approximate $\epsilon_{H}$
as follows

\begin{equation}
\epsilon_{H}=\frac{1/T_{h}^{\alpha_{I}(x_{I},q)}}{1/T_{h}^{\alpha_{I}(x_{I},q)}-1/T_{c}{}^{\alpha_{I}(x_{I},q)}}\approx1-\frac{1}{\alpha_{I}\left(x_{I},q\right)}\frac{T}{\Delta T},\label{eq:91b}
\end{equation}
notice that in the limit of small difference of $\Delta T$,
$\epsilon_{CH}\approx T/\Delta T\approx T_{h}/\Delta T=1/\left(1-\tau\right)$, where $\epsilon_{CH}$ is the Carnot heating COP.

\subsection{Heating dissipation function}

Now, we will calculate the minimum function of heat pump dissipation
for which it is necessary to consider the function of dissipation
that in the case of the pump is given as $\Phi_{IH}=T_{c}\sigma_{I}$,
we can calculate the production of cooling entropy $\sigma_{I}=\left(J_{h}/T_{h}\right)-\left(J_{c}/T_{c}\right)$
which It can be written as follows

\begin{equation}
\sigma_{I}=c_{I}\left(1+\frac{x_{I}}{q}\right)\left[\frac{1}{T_{h}{}^{\alpha_{I}(x_{I},q)+1}}-\frac{1}{T_{c}{}^{\alpha_{I}(x_{I},q)+1}}\right],\label{eq:24-1}
\end{equation}
consider the case of small $\Delta T$,
and Eq.(\ref{ci}) at Eq.(\ref{eq:24-1})

\begin{equation}
\sigma_{I}=-\frac{1}{\epsilon_{C}}\left(\frac{x_{I}^{2}+2qx_{I}+1}{x_{I}^{2}}\right)\frac{L_{22}X_{2}}{T},
\end{equation}
we can calculate the function of heating dissipation Eq.(\ref{eq:4b})

\begin{equation}
\Phi_{IH}=c_{I}T_{c}\left(1+\frac{x_{I}}{q}\right)\left[\frac{1}{T_{h}{}^{\alpha_{I}(x_{I},q)+1}}-\frac{1}{T_{c}{}^{\alpha_{I}(x_{I},q)+1}}\right],\label{eq:25b-2}
\end{equation}
consider the case of small $\Delta T$, substituting Eq.(\ref{td})
and Eq.(\ref{ci}) at Eq.(\ref{eq:25b-2})

\begin{equation}
\Phi_{IH}=-\frac{1}{\epsilon_{C}}\left(\frac{x_{I}^{2}+2qx_{I}+1}{x_{I}^{2}}\right)L_{22}X_{2}.\label{eq:95b}
\end{equation}

\subsection{Generalized Ecological function of heating}

We can define the generalized Ecological function of heating $E_{IGH}$,
in the following manner

\begin{equation}
E_{IG}=J_{h}-g_{E}^{IH}\left(\frac{\epsilon_{MH}}{2}\right)\Phi_{IH},\label{eq:47a-1}
\end{equation}
where $g_{E}^{IH}\left(\epsilon_{MH}/2\right)$ is the function $g_{E}^{IH}$
\cite{Angulo-Brown2001,Arias-Hernandez2003} evaluated at half of
the maximum COP $\epsilon_{MH}$ and $\Phi_{IH}$ is the heating
dissipation function Eq.(\ref{eq:25b-2}), the $g_{E}^{I}$ function
is defined in the following way $g_{E}^{IH}(\epsilon)=\left(\epsilon_{CH}\epsilon\right)/\left(\epsilon_{CH}-\epsilon\right)$,
where $\epsilon_{CH}=1/\left(1-\tau\right)$, the maximun heating
COP $\epsilon_{MH}$ Eq.(\ref{maxcopH}),whereby we can compute $g_{E}^{IH}\left(\epsilon_{MH}/2\right)=\left[\epsilon_{CH}\left(\epsilon_{MH}/2\right)\right]/\left[\epsilon_{CH}-\left(\epsilon_{MH}/2\right)\right]$, with what we obtain

\begin{equation}
g_{E}^{IH}\left(\frac{\epsilon_{MH}}{2}\right)=\frac{\tau^{\alpha_{I}(x_{M\epsilon_{H}},q)}}{2\left[\tau^{\alpha_{I}(x_{M\epsilon_{H}},q)}-1\right]-\frac{1}{\epsilon_{CH}}\left[\tau^{\alpha_{I}(x_{M\epsilon_{H}},q)}\right]},\label{eq:81a-1}
\end{equation}
where $\alpha_{I}\left(x_{M\epsilon_{H}},q\right)$ (see Tab.(\ref{tab:3-1})),
with which we can write $E_{IGH}$ in the following manner 

\begin{equation}
E_{IGH}=c_{I}\left(1+\frac{x_{I}}{q}\right)\left\{ \frac{1}{T_{h}{}^{\alpha}}-g_{E}^{IH}\left(\frac{\epsilon_{MH}}{2}\right)T_{c}\left[\frac{1}{T_{h}{}^{\alpha_{I}(x_{I},q)+1}}-\frac{1}{T_{c}{}^{\alpha_{I}(x_{I},q)+1}}\right]\right\} ,\label{eq:EIG}
\end{equation}
 notice that when $g_{E}^{IH}\left(\epsilon_{MH}/2\right)=1$ then
$E_{IGH}=E_{IH}$, where $E_{IH}$ is the heating ecological function,
consider the case of small temperatures, under this consideration
we can approximate first order Eq.(\ref{eq:81a-1}) as follows

\begin{equation}
g_{E}^{IH}\left(\frac{\epsilon_{MH}}{2}\right)\approx\frac{\epsilon_{MH}\left[\left(1+\sqrt{1-q^{2}}\right)^{2}-\epsilon_{MH}q^{2}\right]}{\left(1+\sqrt{1-q^{2}}\right)^{2}\left(2\epsilon_{MH}-1\right)+\epsilon_{MH}q^{2}},\label{eq:86a-1}
\end{equation}
finally consider the case of small $\Delta T$ , substituting Eq.(\ref{td}),
Eq.(\ref{ci}) and Eq.(\ref{eq:86a-1}) at Eq.(\ref{eq:EIG}) is obtained

\begin{equation}
E_{IGH}=\left(\frac{q+x_{I}}{x_{I}}\right)\left\{ \left[1-\left(\alpha_{I}+1\right)\frac{1}{\epsilon_{C}}\right]+g_{E}^{IH}\left(\frac{\epsilon_{MH}}{2}\right)\left(\alpha_{I}+1\right)\frac{1}{\epsilon_{C}}\right\} L_{22}X_{2}.\label{eq:59a-1-1}
\end{equation}

\subsection{Generalized Omega function of heating}

We can define the generalized Omega function of heating $\Omega_{IGH}$,
this objective function can written as follows

\begin{equation}
\Omega_{IH}=\varepsilon_{e,u}-g_{\Omega}^{I}\left(\frac{\epsilon_{MH}}{2}\right)\varepsilon_{l,u},\label{eq:52a-1-1}
\end{equation}
where $g_{\Omega}^{I}\left(\epsilon_{MH}/2\right)$ is the $g_{\Omega}^{I}\left(\epsilon_{H}\right)$$=\epsilon_{H}/\left(\epsilon_{MH}-\epsilon_{H}\right)$
Omega function  evaluated at half the maximum heating COP
, which is defined as follows $g_{\Omega}^{I}\left(\epsilon_{MH}/2\right)=1$.
For the system operating as a heat pump the useful energy, $\varepsilon_{u}=J_{h}=P+J_{c}$,
the input energy, $\varepsilon_{in}~$, is the work supplied, $P$, the performance coefficient is defined as $r=\epsilon_{H}$
, the minimum performance of the heat is $r_{min}=1$  and the maximum
is $r_{max}=\epsilon_{MH}=1+\epsilon_{M}$ , therefore, if we substitute
this information in the definition of the effective utility energy
$\varepsilon_{e,u}=\varepsilon_{u}-r_{min}\varepsilon_{in}$ , we
arrive at, $\varepsilon_{e,u}=J_{h}-P$. Substituting the maximum
performance of the machine $r_{max}$, the input energy $\varepsilon_{in}\text{}=P$,
and the useful energy $\varepsilon_{u}$, in the definition of the
lost utility energy $\varepsilon_{l,u}=r_{max}\varepsilon_{in}-\varepsilon_{u}$,
we have, $\varepsilon_{l,u}=\epsilon_{MH}P-J_{h}$, substituting $\varepsilon_{u,e}$
and $\varepsilon_{l,u}$ in $\Omega_{IH}=\varepsilon_{u,e}-\varepsilon_{l,u}$
, we obtain $\Omega_{IH}=2J_{h}-\left(1+\epsilon_{MH}\right)P$, with what we obtain

\begin{equation}
\Omega_{IGH}=\left(1+\epsilon_{MH}\right)J_{c}+\left(1-\epsilon_{MH}\right)J_{h},\label{eq:OH1}
\end{equation}

or

\begin{equation}
\Omega_{IGH}=\left(2+\epsilon_{M}\right)J_{c}-\epsilon_{M}J_{h},\label{eq:OH2}
\end{equation}
in the equation Eq.(\ref{eq:OH2}) we have considered $J_{h}=P+J_{c}$
and $r_{max}=\epsilon_{MH}=1+\epsilon_{M}$. Finally, note that Eq.(\ref{eq:OH2})
and Eq.(\ref{eq:87b}) are equal $\Omega_{IH}=\Omega_{IGH}$, this
also happens in the case of small $\Delta T$.

\begin{figure}[tb]
\begin{centering}
a) \includegraphics[scale=0.55]{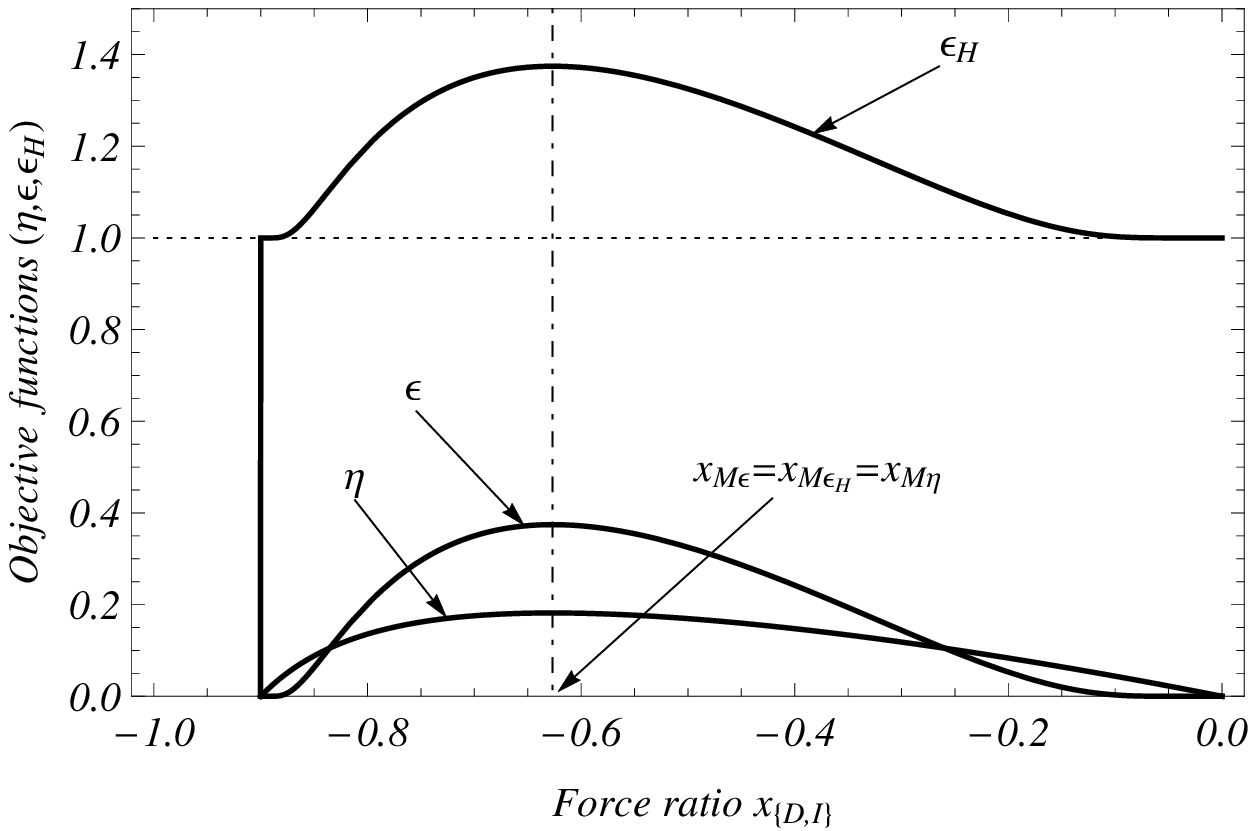} \hspace*{0.35cm}b) \includegraphics[scale=0.34]{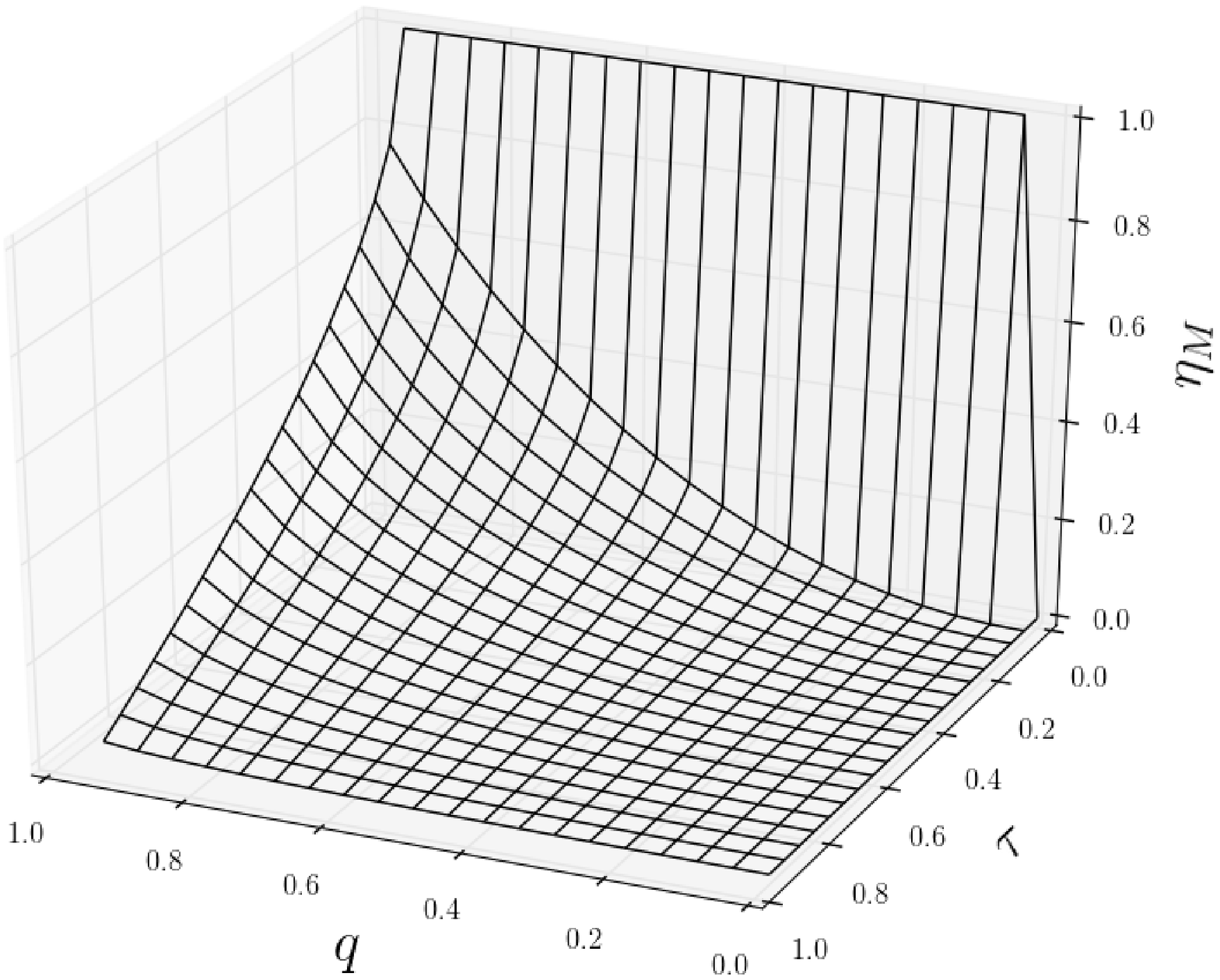}
\par\end{centering}
\begin{centering}
c) \includegraphics[scale=0.34]{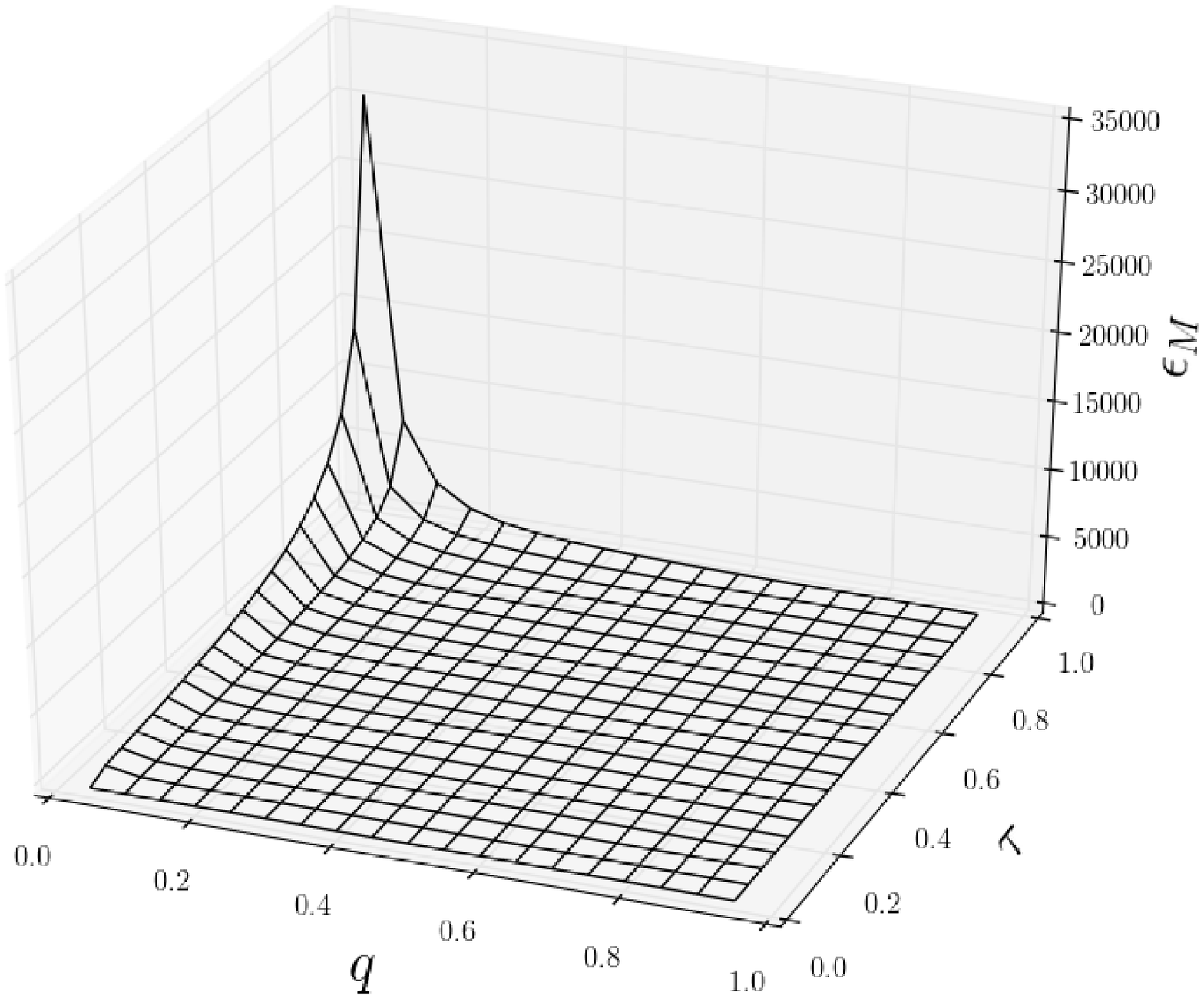}\hspace*{0.35cm}d) \includegraphics[scale=0.34]{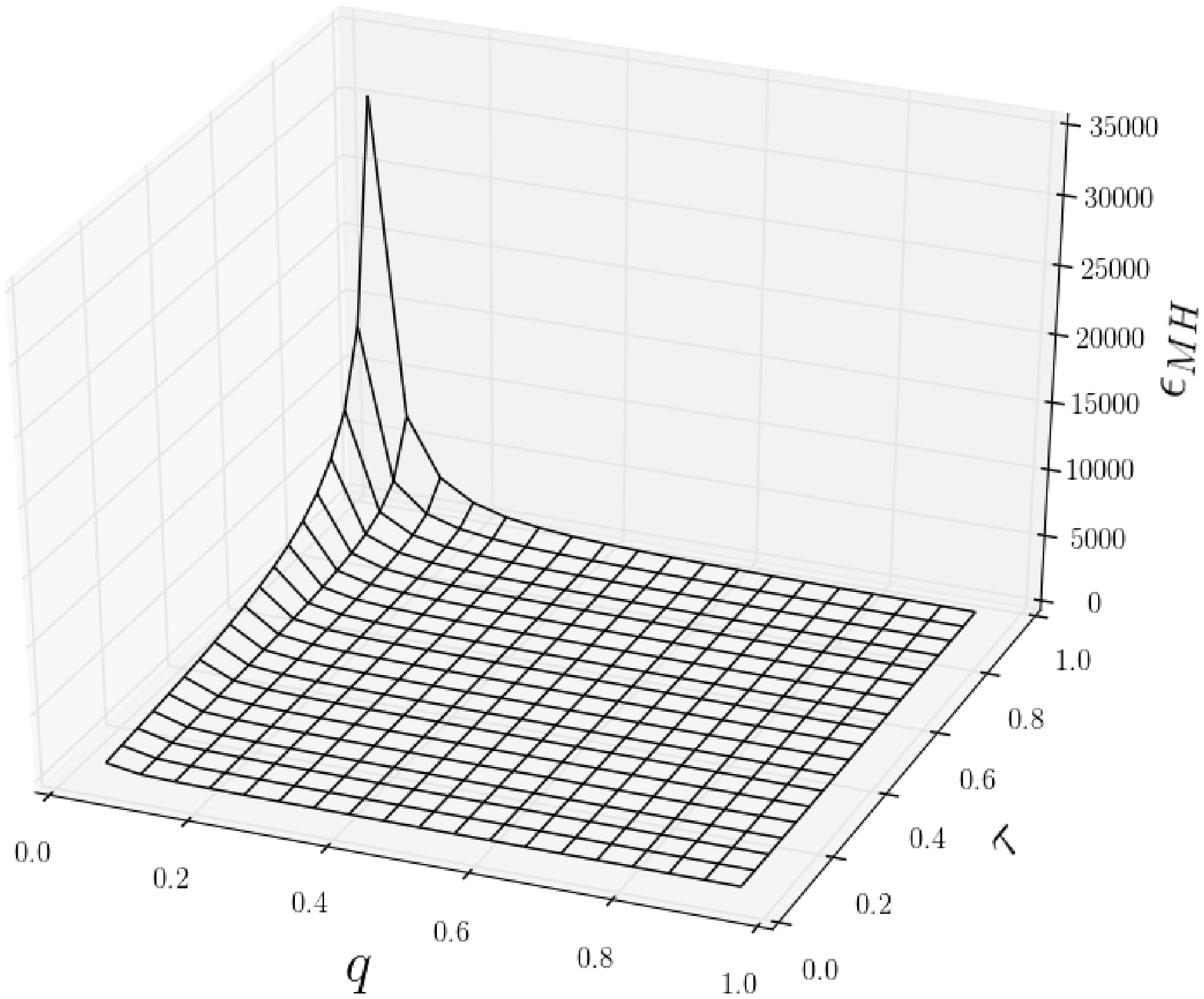}
\par\end{centering}
\begin{quote}
\caption{a) The graph shows the COP $\epsilon$, heating COP $\epsilon_{H}$
and efficiency $\eta$, equations Eq.(\ref{cop}), Eq.(\ref{copH})
and Eq.(\ref{eq:28a}) respectively, with $q=0.9$ and $\tau=0.7$.
As we can see the three operating modes have the same maximum $x_{M\eta}=x_{M\epsilon_{H}}=x_{M\epsilon}$
and that the heating COP satisfy the known relation with the COP,
$\epsilon_{H}=1+\epsilon$. b) In the figure we can see the graph
of the maximum efficiency $\eta_{M\eta}\left(q,\,\tau\right)$ Eq.(\ref{eq:nmaxima}).
c) In the figure we can see the graph of the maximum COP $\epsilon_{M}\left(q,\,\tau\right)$
Eq.(\ref{maxcop}). d) In the figure we can see the graph of the maximum
heating COP $\epsilon_{MH}\left(q,\,\tau\right)$ Eq.(\ref{maxcopH}). }

{\footnotesize{}\label{fig3} }{\footnotesize \par}
\end{quote}
\end{figure}
 
\subsection{Efficient heating power}

We can define the efficient heating power $P_{H\epsilon}$ as the
product of the heating COP $\epsilon_{H}$ by the heating power $P_{H}=J_{h}$,
so that $P_{H\epsilon}=\epsilon_{H}J_{h}$ so that substituting Eq.(\ref{perj})
and Eq.(\ref{copH}) in our definition of $P_{H\epsilon}$ we obtain

\begin{equation}
P_{H\epsilon}=c_{I}\left(1+\frac{x_{I}}{q}\right)\frac{T_{c}{}^{\alpha_{I}(x_{I},q)}}{T_{h}{}^{\alpha_{I}(x_{I},q)}\left[T_{c}{}^{\alpha_{I}(x_{I},q)}-T_{h}{}^{\alpha_{I}(x_{I},q)}\right]},\label{eq:96b}
\end{equation}
consider the case of small $\Delta T$

\begin{equation}
\frac{T_{c}{}^{\alpha_{I}(x_{I},q)}}{T_{h}{}^{\alpha_{I}(x_{I},q)}\left[T_{c}{}^{\alpha_{I}(x_{I},q)}-T_{h}{}^{\alpha_{I}(x_{I},q)}\right]}\approx-\frac{1}{\alpha_{I}\left(x_{I},q\right)}\frac{1}{T_{h}{}^{\alpha_{I}(x_{I},q)}}\frac{T}{\varDelta T}=-\frac{1}{\alpha_{I}\left(x_{I},q\right)}\frac{1}{T_{h}{}^{\alpha_{I}(x_{I},q)}}\epsilon_{C},
\end{equation}
substituting Eq.(\ref{td}), Eq.(\ref{ci}) and Eq.(\ref{eq:91b})
at Eq.(\ref{eq:96b}), we obtain

\begin{equation}
P_{H\epsilon}=-\frac{\left(x_{I}+q\right)^{2}}{qx_{I}+1}\epsilon_{C}L_{22}X_{2},
\end{equation}
this function does not have a point of interest for the optimization of the arrangement, but this function does have it for non-linear systems.

\section{Concluding remarks\label{conclusions}}

In this paper, we presented a proposal to obtain the temperature profile for an assemblage of linear energy converters (machine--elements) as published by  \cite{Jimenez2007,Jimenez2008}, which in principle yields a Riccati's equation for that profile. In order to avoid solving the above differential equation, we used the generalized fluxes of a non--isothermal converter, under the Onsager description, written in terms of the force ratio $x_{i}$ and the coupling coefficient $q$. With this scheme it is possible to calculate the fluxes of heat $J_{a}$ and $J_{b}$, which are completely determinate except for an integration constant $c_{0_{i}}$. This integration constant is related to an initial condition of the temperature profile $f\left(y\right)$. In this work, we find this constant of integration considering the case of the perfect coupling $q=1$ and under the assumption of small temperature differences.

In addition to obtaining the profile $f(y)$ and therefore the fluxes $J_{a}$ and $J_{b}$, with these fluxes it is possible to build the output power and the entropy production, in addition to different objective functions which we use to study the energetics of the assemblage operating as D--LEC or I--LEC. 
In the same way, we determine the condition to approximate these objective functions to describe the energetics of a non--isothermal single lineal energy converter.

On the other hand, as we can see in the figure Fig. (\ref{fig3})  the efficiency $\eta$ (\ref{eq:28a}), the COP $\epsilon$ (\ref{cop})
and the heating COP $\epsilon_{H}$ (\ref{copH}) have the same optimum value of the force ratio  $x_{M\eta}=x_{M\epsilon}=x_{M\epsilon_{H}}$. This optimum value of the  force ratio is independent of any initial condition imposed on the assemblage of machine--elements (heat law), this is a very interesting result since in principle the objective of the energy converter is different (although the three measure the performance in the conversion of energy), but in spite of these differences they are subject to the second law of thermodynamics.

\subsection*{Acknowledgement}

We thank Dr. Fernando Angulo Brown for stimulating discussions, suggestions and invaluable
help in the preparation of the manuscript. This work was supported
by CONACYT, M\'exico.


\begin{thebibliography}{100}

\bibitem{Jimenez2007}B. Jim\'enez de Cisneros and A. Calvo Hernandez,
Collective Working Regimes for Coupled Heat Engines, Phys. Rev. Lett. \textbf{98}, 130602 (2007)

\bibitem{Jimenez2008}B. Jim\'enez de Cisneros and A. Calvo Hernandez, Coupled heat devices in linear irreversible thermodynamics, Phys. Rev E. \textbf{77}, 041127 (2008).

\bibitem{STUCKI1980} J. W. Stucki, The Optimal Efficiency and the
Economic Degrees of Coupling of Oxidative Phosphorylation, European
Journal of Biochemistry. \textbf{109}, 269 (1980).

\bibitem{van05} C. Van den Broeck, Thermodynamic Efficiency at Maximum Power, Phys. Rev. Lett. \textbf{95}, 190602 (2005).

\bibitem{jimenez06}B. Jim\'enez de Cisneros, L. A. Arias-Hernandez 
and Hernandez, A. Calvo, Linear irreversible thermodynamics and coefficient
of performance, Phys. Rev. E.\textbf{73}, 057103 (2006).

\bibitem{Elsgoltz}L. Elsgoltz, Ecuaciones diferenciales y c\'alculo
variacional, Editorial MIR, Tercera edicion, (1983).

\bibitem{Caplan1983}S. R. Caplan and A. Essig, Bioenergetics
and Linear Nonequilibrium Thermodynamics: the Steady State , 1st ed.(
Cambridge, MA: Ed. Harvad University Press, 1983).

\bibitem{onsager31a} L. Onsager, Reciprocal Relations
in Irreversible Processes. I, Phys. Rev. \textbf{37}, 405 (1931).

\bibitem{onsager31b}L. Onsager, Reciprocal Relations
in Irreversible Processes. II, Phys. Rev. \textbf{38}, 382265 (1931).

\bibitem{Arias-Hernandez2008} L. A. Arias-Hernandez, F. Angulo-Brown,
 and R. T. Paez-Hernandez, First-order irreversible thermodynamic
approach to a simple energy converter, Phys. Rev. E. \textbf{77}, 011123 (2008).

\bibitem{Yourgrau}W. Yourgrau, A. van der Merwe, Gough Raw,
Treatise on Irreversible and Statistical Thermodynamics: An Introduction
to Nonclassical Thermodynamics, (Dover Pubns, Dover Books on
Physics, 2002).

\bibitem{degroot84}S. R. de Groot, P Mazur, Non-equilibrium thermodynamics
(Dover Publications, 1984).

\bibitem{carnot}S. Carnot, Physics, Reflections
on the Motive Power of Fire: And Other Papers on the Second Law of
Thermodynamics, (Dover Publications 2005).

\bibitem{CA}F. L. Curzon and B. Ahlborn, Efficiency of a Carnot engine at maximum power output, Am. J. Phys. \textbf{43}, 22 (1975)

\bibitem{Yilmaz}T. Yilmaz, A new performance criterion for heat engines:
efficient power, Journal of the Energy Institute. \textbf{79}, 38 (2006).

\bibitem{tribus61}Tribus Myron, Thermostatistics and thermodynamics
(D.Van Nostrand Company, Inc., 1961).

\bibitem{tornez06}L. Partido Tornez, Aplicaci\'on del criterio
omega y ecol\'ogico generalizados a diferentes convertidores de energ\'{\i}a,
Tesis de Maestr\'{\i}a, ESFM-IPN, M\'exico (2006).

\bibitem{Angulo-Brown2001}F. Angulo\textendash Brown and L. A. Arias\textendash Hernandez, Reply to Comment on: A general property of endoreversible
thermal engines, Journal of Applied Physics. \textbf{81}, 1520 (2001).

\bibitem{Arias-Hernandez2003} L. A. Arias-Hernandez, G. Ares de Parga,
 and  F. Angulo-Brown, On some nonendoreversible engine models
with nonlinear heat transfer laws, Open Systems \& Information
Dynamics. \textbf{10}, 351 (2003).

\bibitem{angulo91}F. Angul--Brown, An ecological optimization criterion
for finite-time heat engines, J. Appl. Phys. \textbf{69}, 7465 (1991).

\bibitem{calvo01}A. Calvo-Hernandez, A. Medina, J. M. M. Roco, J.
A White and S. Velasco, Unified optimization criterion for energy
converters. Phys. Rev. E. \textbf{63}, 037102 (2001).
\end{thebibliography}
\end{document}